\documentclass{aa}

\usepackage{graphicx, psfig, times, amsfonts}

\newcommand{\less}{\raisebox{-1.1mm}{$\stackrel{<}{\sim}$}}
\newcommand{\more}{\raisebox{-1.1mm}{$\stackrel{>}{\sim}$}}

\newcommand{\msolyr}{{M$_{\odot}$}yr$^{-1}$}

\newcommand{\lsol}{{L$_{\odot}$}}

\newcommand{\ks}{km s$^{-1}$}
\newcommand{\mdot}{$\dot{M}$}

\begin{document}



\title{The mid- and far-infrared colours of AGB and post-AGB 
stars\thanks{Appendix~A is available in the on-line edition of A\&A.}
}

\author{M.A.T. Groenewegen
}

\offprints{Martin Groenewegen (groen@ster.kuleuven.be)}

\institute{
Instituut voor Sterrenkunde, Katholieke Universiteit Leuven, Celestijnenlaan 200B, 
B-3001 Leuven, Belgium
}

\date{received,  accepted: 27 October 2005}

\abstract { 
With the advent of space missions, like SPITZER and ASTRO-F, with
sensitive detectors in the near- and mid-infra red covering a reasonable 
field-of-view and having a good spatial resolution, it will be possible
to detect individual AGB stars in Local Group galaxies. The filters used by these
missions are non-standard and different from mission to mission.
In this paper, the colours of mass-losing AGB and post-AGB stars are
calculated in the broad-band filters of the SPITZER and ASTRO-F
missions, as well as Bessell $V,I$ and 2MASS $J,H,K$ to connect these
results to existing ground-based data.
The models are calculated for carbon- and oxygen-rich chemistry and
cover different effective temperatures and dust compositions.
}

\maketitle

\section{Introduction}

Asymptotic Giant Branch (AGB) and post-AGB stars are prominent
emitters in the infrared (IR), firstly because of their low effective
temperatures and secondly because of their mass loss rates, which can
go up to 10$^{-4}$ \msolyr\ for brief spells of time, in conjuction
with dust formation. The importance of the mid-IR was first
demonstrated by the results of the IRAS mission (Beichman et al. 1988)
which discovered many mass-losing AGB stars (e.g. van de Veen \&
Habing 1988) in the solar neighbourhood but also luminous
mass-losing AGB stars in the Large Magellanic Cloud (LMC) (see Reid et
al. 1990, Reid 1991, Wood et al. 1992, Zijlstra et al. 1996, Loup et
al. 1997), and Small MC (Whitelock et al. 1989, Groenewegen \&
Blommaert 1998), at the limiting magnitudes of IRAS near 100 mJy at 12
micron.  When the ISO satellite was launched many of these mass-losing
AGB stars discovered by IRAS in the MCs were followed up (e.g. Trams
et al. 1999, van Loon et al. 1999), and new surveys in the
MCs were conducted (Loup et al. 1999a, b), in particular using the
ISOCAM instrument (Cesarsky et al. 1996, Blommaert et al. 2003) that
had imaging filters covering 3 to 15 micron with a field-of-view of
typically 1.5\arcmin\ by 1.5\arcmin.  In addition, the Galactic Centre
region, where IRAS was severely confusion limited, was imaged during
the ISOGAL survey (Omont et al. 2003), discovering many AGB stars
(Omont et al. 1999).

IRAS and ISO showed the importance of the IR region, from the Near-IR
to the Far-IR.  The SPITZER mission (Werner et al. 2004), launched on
25 August 2003, and the ASTRO-F mission (e.g. Pearson et al. 2004,
Matsuhura et al. 2005), with a planned launch date of spring 2006,
carry a range of imaging filters that cover the near-, mid- and
far-IR. Due to an improved sensitivity and spatial resolution it is
expected to resolve individual AGB stars in galaxies far beyond the
MCs.

In the present work, series of dust radiative transfer models are
presented that cover the spectral types, dust composition and mass
loss rates seen for Galactic AGB stars. The flux-densities are
computed in the SPITZER IRAC (Fazio et al. 2004) and MIPS (Rieke et
al. 2004) filters, and the ASTRO-F IRC, MIR-S, MIR-L and FIS filters. For
comparison to ground based data $V,I$ and $J,H,K$ magnitudes are also
presented.

Sect.~2 presents the radiative transfer model and the inputs to it,
including the various dust compositions considered. Sect.~3 gives
details on the considered filter curves and the adopted
flux-calibration. Sect.~4 presents the results, scaling laws to
arbitrary luminosities and distances, and some caveats in the use of
the results. Sect.~5 concludes by illustrating colour-magnitude
diagrams for AGB stars in M31 and the Wolf-Lundmark-Melotte (WLM) galaxy.

Earlier results from this kind of dust modelling (in particular for
the ISOCAM filters) were available as private communication and have
been used in Blommaert et al. (2000), Ortiz et al. (2002) and Ojha et
al. (2003).

\section{Inputs to the Radiative Transfer model}

The models have been calculated with a 1-dimensional dust radiative
transfer (RT) code that solves the radiative transfer equation and the
thermal balance equation in a self-consistent way (Groenewegen 1993,
also see Groenewegen 1995).

Basic inputs to the model are the stellar luminosity ($L$), distance ($d$),
photospheric spectrum, (total) mass loss rate, dust-to-gas ratio,
(terminal) outflow velocity, dust condensation temperature ($T_{\rm c}$) 
and composition of the dust.

The models have been calculated for (arbitrary) values of $L$ = 3000 \lsol, 
$d$ = 8.5 kpc, $v_{\infty}$ = 10 \ks, dust-to-gas ($\Psi$) ratio = 0.005,
and no interstellar reddening. 
Scaling relations will be presented later.

Photospheric input spectrum for O-rich stars are taken from Fluks et
al. (1994) for spectral types M0 ($T_{\rm eff}$ = 3850 K), 
M6 ($T_{\rm eff}$ = 3297 K) and M10 ($T_{\rm eff}$ = 2500 K). For C-rich 
stars the models by Loidl et al. (2001) for $T_{\rm eff}$ = 3600 and 
2650 K are considered\footnote{And with a C/O ratio of 1.1.}.

\begin{figure}
\includegraphics[width=85mm]{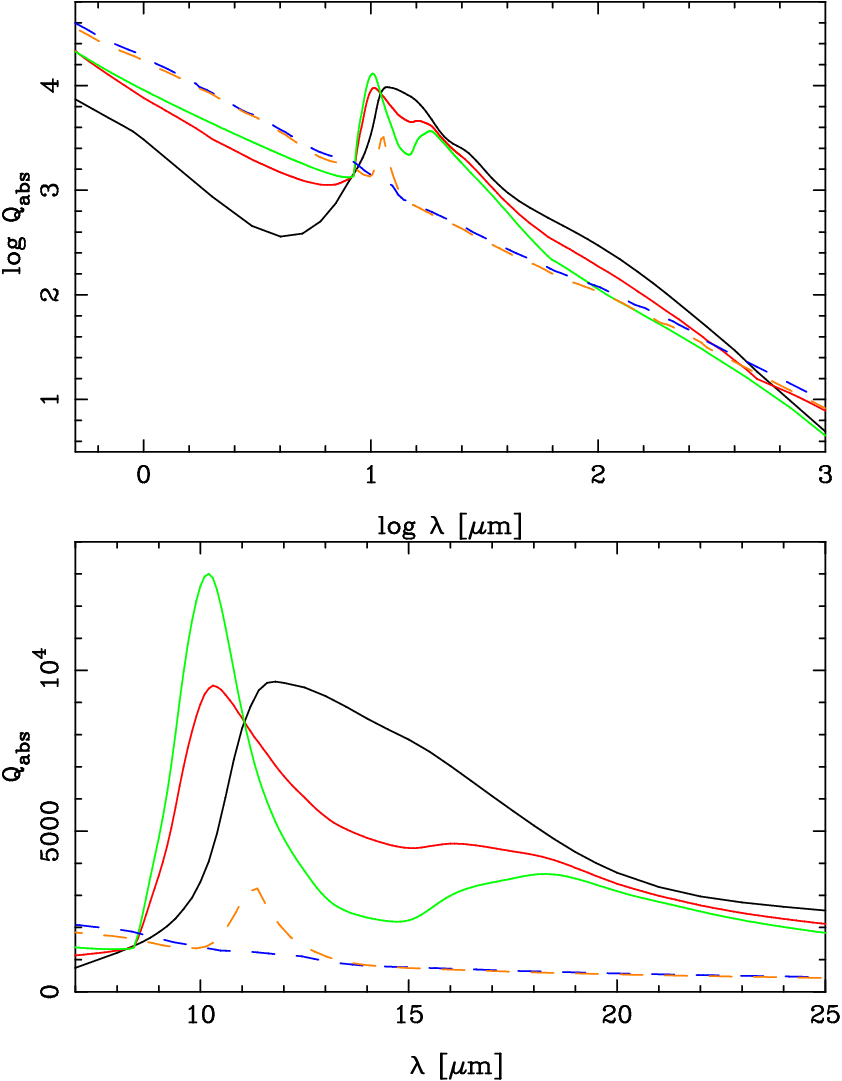}
\caption[]{
The absorption efficiency for the five dust species considered here,
plotted over the whole wavelength range (top panel), and the
mid-IR. Dashed lines indicate 100\% AMC, and the combination of 85\%
AMC + 15\% SiC, which peaks near 11.3 $\mu$m. The solid lines indicate
100\% silicate (with a narrow feature that peaks near 10 $\mu$m), 40\%
silicate + 60\% AlOx, and 100\% AlOx (with a broad feature that peaks
near 12 $\mu$m).
}
\label{Fig-Dust}
\end{figure}

Several types of dust are considered that cover the main features
observed in AGB stars.  The absorption coefficients as a function of
wavelength are displayed in Figure~\ref{Fig-Dust}. For dust around
O-rich stars they are:

\begin{itemize}

\item
100\% Aluminium Oxide (AlOx; amorphous porous Al$_2$O$_3$), with
optical constants from Begemann et al. (1997) and assuming a
condensation temperature of $T_{\rm c} = 1500$ K. 

\item
A combination of 60\% AlOx and 40\% Silicate (optical constants from
David \& Pegourie 1995) and assuming a condensation temperature of $T_{\rm c} = 1500$ K. \\ 

\noindent
The dust species with 60 to 100\% AlOx can explain the observed
Spectral Energy Distributions (SEDs) and ISOCAM CVF 5-14 $\mu$m
spectra in Galactic Bulge AGB stars which have mass loss rates up to
10$^{-7}$ \msolyr\ (Blommaert 2003, Blommaert et al. 2005, in preparation).

\item
100\% silicate with $T_{\rm c} = 1000$ K.

\end{itemize}

\noindent
For dust around C-rich stars two species are considered:

\begin{itemize}

\item
A combination of 85\% Amorphous Carbon (AMC) and 15\% Silicon Carbide
(SiC) with optical constants from, respectively, Rouleau \& Martin
(1991) for the AC1 species and $\alpha$-SiC from P\'{e}gouri\'{e} (1988). 
A $T_{\rm c}$ of 1200 K is adopted.

\item
100\% AMC with a  $T_{\rm c}$ of 1000 K.

\end{itemize}

\noindent
This range of AMC with zero to 15\% percent SiC and the corresponding
condensation temperatures can explain the SEDs and IRAS LRS spectra of
the majority of (Galactic) C-stars (Groenewegen 1995, Groenewegen et al. 1998).
For all dust species a uniform grain size of 0.1 $\mu$m is adopted.

\section{Flux Calibration}

\begin{figure*}
\includegraphics[width=170mm]{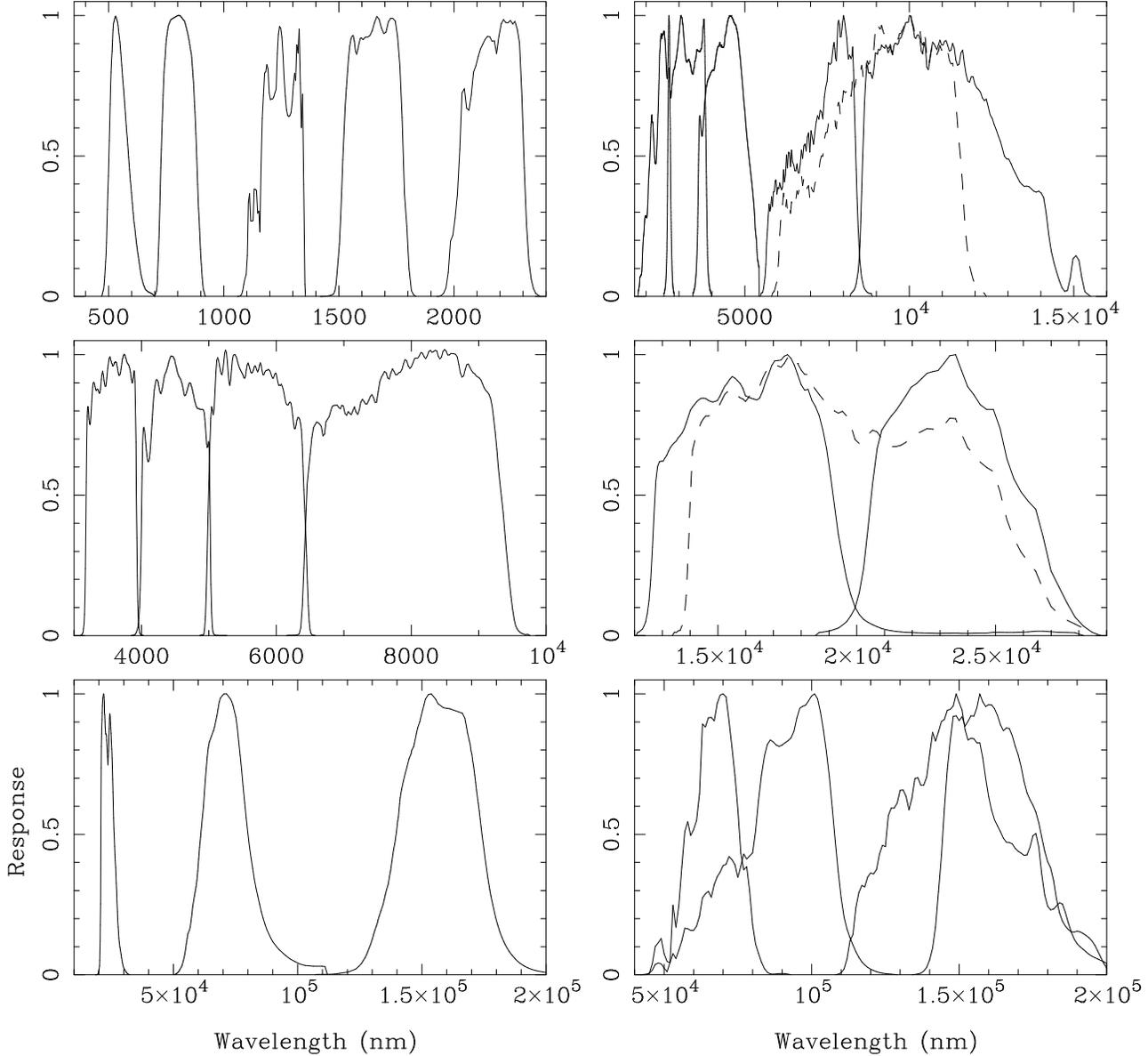}
\caption[]{
Filter curves adopted: Bessell $VI, 2MASS JHK$ (left top), Spitzer-IRAC (left
middle), Spitzer-MIPS (left bottom), ASTRO-F IRC and MIR-S (right
top), ASTRO-F MIR-L (right middle), and ASTRO-F FIS (right bottom).
}
\label{Fig-Filters}
\end{figure*}

In this paper the broad-band filters of the SPITZER and ASTRO-F
missions are considered. Both contain filters in the near- and mid-IR
which are ideal to identify mass losing stars. Filtercurves have been
obtained from official
websites\footnote{http://ssc.spitzer.caltech.edu/mips/spectral\_response.html;
http://ssc.spitzer.caltech.edu/irac/spectral\_response.html;
http://www.ir.isas.jaxa.jp/ASTRO-F/Observation/}, and are reproduced
in Fig.~\ref{Fig-Filters}.  The filter names and their central
wavelengths are given in Table~\ref{tab-vega}.

To connect this data to existing ground-based data Bessell $V,I$ and
2MASS $J,H,K$ magnitudes are also calculated, whose filter curves are
also shown in Fig.~\ref{Fig-Filters}.

The main output of the RT code is the flux emerging from the AGB star
and its dust envelope. This flux is folded with the filter curves and
flux-densities and magnitudes are calculated from (following the
definition used by IRAS, ISOCAM and IRAC, see e.g. Blommaert et al 2003):

\begin{equation}
m_\lambda = -2.5 \log \left( \frac{\int  (\lambda/\lambda_0) \; F_\lambda \; R_\lambda \;
d\lambda}{\int R_\lambda \; d\lambda} \right) + m_0
\end{equation}

The zero-points, $m_0$, are calculated from a reference spectrum for
Vega calculated using the MARCS code (Decin, private communication).
The resulting and adopted flux-densities for a zero-magnitude star are
listed in Table~\ref{tab-vega}.

\begin{table}
\caption{Adopted flux-densities for a zero-magnitude star}
\begin{tabular}{lrr} \hline

Filter      & $\lambda_0$ ($\mu$m) & Flux (Jy)  \\ \hline
IRAC\_3.6  & 3.55 &  278.8 \\
IRAC\_4.5  & 4.49 &  179.6 \\
IRAC\_5.8  & 5.73 &  118.4 \\
IRAC\_8.0  & 7.87 &   63.5 \\
MIPS\_24   & 24    &   7.23 \\
MIPS\_70   & 70    &   0.791 \\
MIPS\_160  & 160   &   0.190 \\
IRC\_N2    &  2.69 & 651.4 \\
IRC\_N3    &  3.06 & 323.5 \\
IRC\_N4    &  4.55 & 199.8 \\
MIRS\_S7   &  8.0  &  84.5 \\
MIRS\_S9W  & 10.04 &  57.7 \\
MIRS\_S11  & 10.0  &  32.0 \\
MIRL\_L15  & 17.5  &  17.6 \\
MIRL\_L18W & 17.6  &  11.0 \\
MIRL\_L24  & 23.55 &   7.45 \\
FIS\_N60   & 60    &   0.935 \\
FIS\_WIDE-S & 90   &   0.604 \\
FIS\_WIDE-L & 140  &   0.194 \\
FIS\_N160   & 160  &   0.166 \\

\hline
\end{tabular}
\label{tab-vega}
\end{table}

\section{Results for AGB and post-AGB stars}

\begin{figure}
\includegraphics[angle=-90,width=85mm]{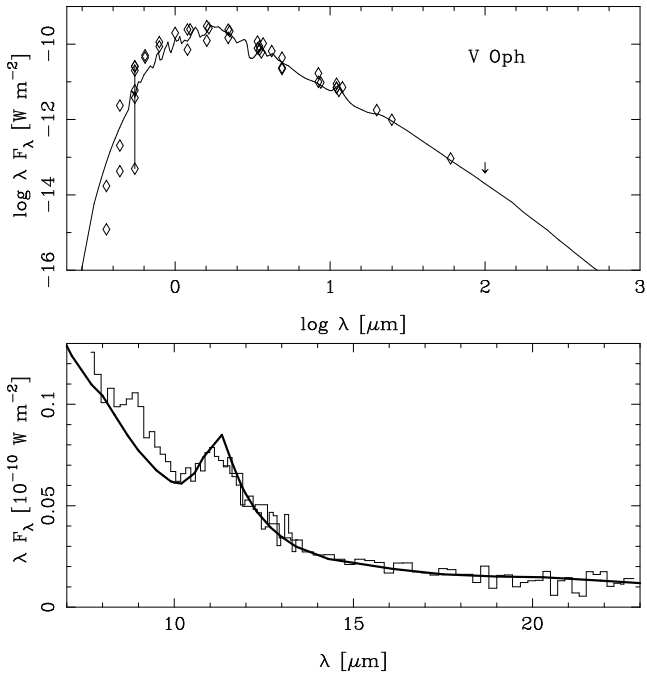}
\caption[]{
Model fit to the SED and IRAS LRS spectrum of the carbon Mira V Oph
(cf. Groenewegen et al. 1998).  Used is the 2650 K model atmosphere,
$L$ = 4390 \lsol, $d$ = 0.68 kpc, \mdot = 3.0 $\times 10^{-8}$ \msolyr, and a
combination of 85\% AMC and 15\% SiC.
}
\label{fig-voph}
\end{figure}

As mentioned before, the results of the numerical code have already
been used to succesfully fit the SEDs of individual stars (Groenewegen
1994a,b, 1995, 1997; Groenewegen et al. 1995, 1997, 1998) and predict
the colours of AGB stars (Blommaert et al. 2000, Ortiz et al. 2002,
Ojha et al. 2003) and has been bench-marked against other codes (Ivezic
et al. 1997). To illustrate the typical accuracies that can be achieved
in fitting SEDs and colours, Figure~\ref{fig-voph} shows a fit to the
SED and the IRAS LRS spectrum of the Mira V Oph (cf. Figure~33 in
Groenewegen et al. 1998), but now with the 2650 K model atmosphere
instead of a black-body in Groenewegen et al. (1998). The agreement is
quite good over the whole wavelength range.

\begin{figure}
\includegraphics[width=85mm]{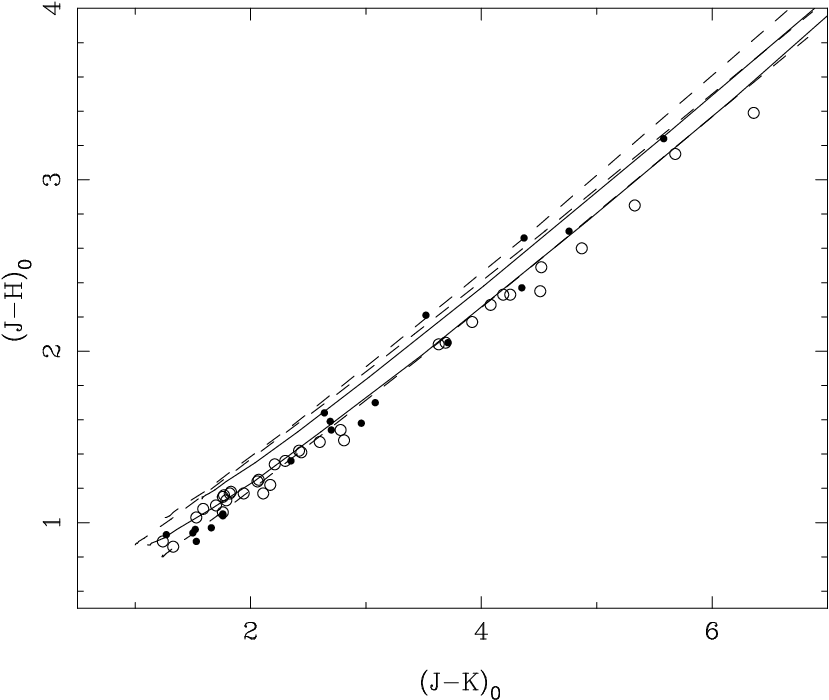}
\includegraphics[width=85mm]{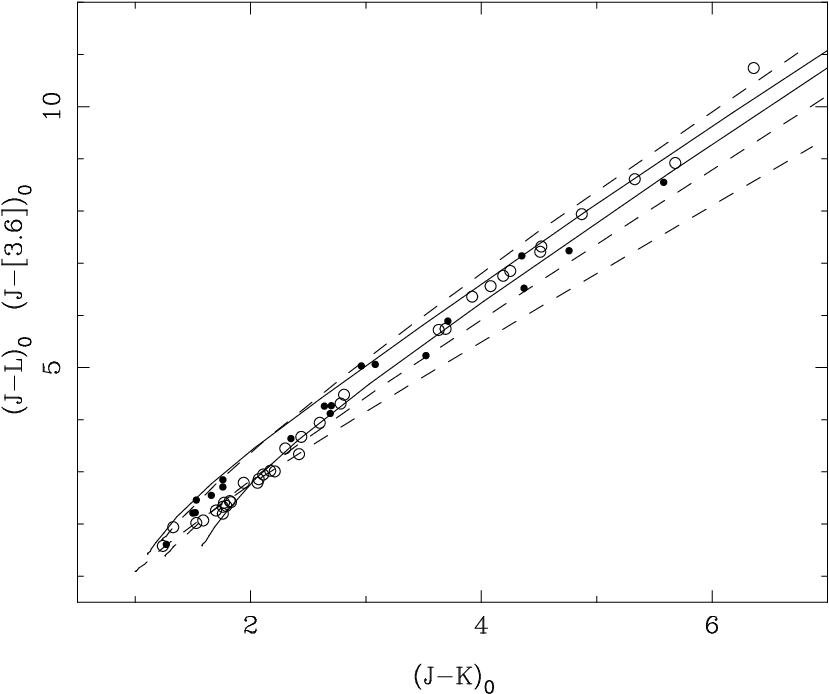}
\includegraphics[width=85mm]{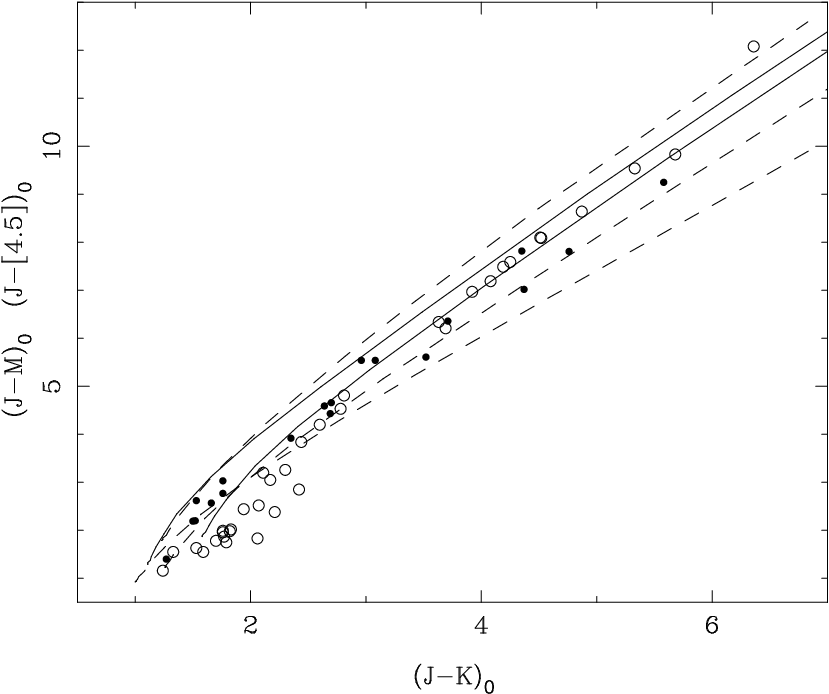}

\caption[]{
Colour-Colour diagrams with $(J-K)$ plotted against $(J-H)$,
$(J-[3.6])$ ($(J-L)$ for the observations), and $(J-[4.5])$ ($(J-M)$ for
the observations).
Carbon stars are plotted as open circles, oxygen-rich stars as filled circles.
Shown are model sequences for carbon stars (solid lines) for
$T_{\rm eff}$ = 3600 K, 85\% AMC + 15\% SiC dust, and 
$T_{\rm eff}$ = 2650 K, 100\% AMC dust (the redder sequence), and
oxygen-rich stars (dashed lines) for spectral type M0 and 100\% AlOx dust, 
M6 and 40\% silicate + 60\% AlOx, and M10 and 100\% silicate dust 
(from top to bottom for a given $(J-K)$).
}
\label{fig-cc}
\end{figure}

Figure~\ref{fig-cc} shows colour-colour diagrams of $(J-K)$ versus
$(J-H)$, $(J-[3.6])$ and $(J-[4.5])$ (the latter two being
representative for $(J-L)$ and $(J-M)$, respectively). The data are
the mean $JHKLM$ magnitudes of carbon and oxygen-rich stars based on the
monitoring of the light curves from Le Bertre (1992, 1993) and
Taranova \& Shenavrin (2004). Only stars with magnitudes in all bands
and without uncertain magnitudes were considered. 

Although such a comparison between models and data is not trivial
because of the effects of reddening (although most sources are within
2.5 kpc and reddening at these wavelengths should be small) and the
fact that the IRAC [3.6] and [4.5] filters are not directly comparable
to $L$ and $M$, it may nevertheless illustrate the overall correctness
of the models, as suggested by the referee. Again, the overall
agreement between models and data is quite good with most data points
within 0.1 magnitude of any of the sequences.

The results of the models obtained in the present paper for the $V, I,
J, H, K$, IRAC, MIPS and ASTRO-F magnitudes and fluxes are listed in
Tables~A.1 to A.26 for AGB stars and Tables~A.27 to A.36 for post-AGB stars. 
Table~A.1 is repeated as Table~2 to illustrate the content.

The AGB models have been calculated for mass loss rates up to a value
resulting in a $(J-K)$ colour of about 10, roughly the reddest known
AGB stars (Groenewegen et al. 1998).
For every set of input in terms of stellar input spectrum and dust
composition there are two tables, listing, respectively, $V,I,J,H,K$
(in magnitudes) and the flux-densities (in mJy) for SPITZER IRAC and
MIPS, and the flux-densities (in mJy) for the ASTRO-F IRC, MIR-S,
MIR-L and FIS filters. Tables~A.1-A.26 also list the mass loss rate
and the dust optical depth, at
11.75 $\mu$m for AMC+SiC and 
11.33 $\mu$m for AMC,
11.75 $\mu$m for AlOx and AlOx+silicate, and
10.20 $\mu$m for silicate.
Optical depths at other wavelengths can be estimated from Fig.~\ref{Fig-Dust}.
For the post-AGB models the dust temperature at the inner radius is
listed instead of the mass loss rate. By decreasing this parameter the
expansion of the dust envelope after cessation of mass loss at the end
of the AGB can be simulated. 
The post-AGB tracks have only been calculated for the 100\% silicate
and 100\% AMC models, as it appears that dust with appreciable amounts
of, respectively, AlOx and SiC only appear at lower mass loss rates,
and not at the tip of the AGB.
The model runs are such that the first entry for the post-AGB models
corresponds to the same parameter set as for the last entry in the
corresponding AGB model (e.g. Tab.~A.27-A.3, A.29-A.11, A.34-A.24) but
listed as the result of an independent calculation. One can notice
small differences of $\less$1\% in the fluxes and this is due to the
accuracy of the RT model at these large optical depths.

\subsection{Scaling laws}

The fluxes listed in the tables scale with $(L/3000) / (d/ 8.5)^2$
(and the magnitudes as $-2.5 \log$ of this factor).

The spectral energy distribution is--for a given stellar input 
spectrum--determined only by the dust optical depth, defined by (Groenewegen 1993, 1995):
\begin{displaymath}
{\tau}_{\lambda} = \int_{r_{inner}}^{r_{outer}} \pi a^2 Q_{\lambda} \; n_d(r) 
\; dr  = 
\end{displaymath}
\begin{equation}
= 5.405 \; 10^8 \; \frac{\dot{M}  \; \Psi \; Q_{\lambda}/a}{R_{\star} \; 
v_{\infty} \; {\rho}_d \; r_c} \; \int_{1}^{x_{max}} \frac{R(x)}{x^2}\; dx
\end{equation}
where $n_d$ is the number density of dust particles, $x = r/r_{\rm c}$
and $\dot{M}$(r) = $\dot{M} \, R(x)$. The units are: the (present-day)
mass loss rate at the inner radius $\dot{M}$ in \msolyr, $\Psi$ the
dust-to-gas mass ratio, $Q_{\lambda}/a$ the absorption coefficient of
the dust over the grain radius in cm$^{-1}$, $R_{\star}$ the stellar
radius in solar radii, $v_{\infty}$ the terminal velocity of the
circumstellar envelope in km s$^{-1}$, ${\rho}_d$ the dust grain
density in gr cm$^{-3}$, $r_{\rm c}$ the inner dust radius in stellar
radii and $x_{\rm max}$ the outer radius in units of $r_{\rm c}$. For
the assumptions adopted in the present paper--a constant velocity and
mass loss rate, and a outer radius much larger than the inner
radius--the integral becomes unity.

This relation implies a necessary scaling when the expansion velocity
and/or dust-to-gas ratio and/or luminosity are different from the nominal ones, and this
scaling is like \mdot $\sim (v/10\;$ \ks) (0.005/$\Psi$) $\sqrt{L/3000}$.

\begin{table*}
\caption{Optical, NIR, IRAC and MIPS magnitudes and fluxes for a
C-rich AGB star, $T_{\rm eff}$ = 2650 K, 85\% AMC + 15\% SiC.  All
tables are available on-line in Appendix~A as Table~A.1-A.36.  In all
tables \mdot\ represents the mass loss rate in \msolyr, while $\tau$
represents the dust optical depth at the wavelengths listed in
Sect.~4.}

\begin{tabular}{rrrrrrrrrrrrrr} \hline
\small

\mdot\   & $\tau$    &  $V$  &  $I$  &  $J$  &  $H$  &  $K$  &    3.6 &    4.5 &    5.8 &    8.0 &   24    &   70    & 160  \\ \hline

0.10E-09 &  0.000045 & 13.85 & 10.99 &  9.07 &  7.92 &  7.49 &  282.3 &  236.6 &  140.1 &   69.2 &   23.36 &    2.91 &   0.773 \\  
0.25E-09 &  0.000113 & 13.85 & 10.99 &  9.07 &  7.92 &  7.49 &  282.5 &  236.9 &  140.3 &   69.4 &   23.42 &    2.92 &   0.776 \\  
0.50E-09 &  0.000226 & 13.85 & 10.99 &  9.07 &  7.92 &  7.49 &  282.9 &  237.3 &  140.7 &   69.7 &   23.51 &    2.93 &   0.781 \\  
0.10E-08 &  0.000452 & 13.85 & 11.00 &  9.07 &  7.92 &  7.49 &  283.6 &  238.1 &  141.5 &   70.5 &   23.71 &    2.97 &   0.790 \\  
0.25E-08 &  0.001130 & 13.86 & 11.00 &  9.07 &  7.93 &  7.49 &  285.9 &  240.5 &  143.8 &   72.6 &   24.28 &    3.08 &   0.817 \\  
0.50E-08 &  0.002259 & 13.87 & 11.01 &  9.08 &  7.93 &  7.49 &  289.7 &  244.6 &  147.7 &   76.1 &   25.24 &    3.26 &   0.863 \\  
0.10E-07 &  0.004513 & 13.89 & 11.02 &  9.08 &  7.93 &  7.49 &  297.2 &  252.7 &  155.4 &   83.1 &   27.14 &    3.62 &   0.953 \\  
0.25E-07 &  0.011247 & 13.96 & 11.07 &  9.11 &  7.95 &  7.48 &  319.2 &  276.5 &  178.3 &  103.7 &   32.77 &    4.69 &   1.223 \\  
0.50E-07 &  0.022382 & 14.08 & 11.15 &  9.16 &  7.98 &  7.48 &  354.4 &  314.8 &  215.1 &  137.2 &   41.92 &    6.43 &   1.663 \\  
0.10E-06 &  0.044342 & 14.31 & 11.31 &  9.25 &  8.03 &  7.49 &  418.4 &  385.8 &  284.2 &  200.4 &   59.44 &    9.75 &   2.498 \\  
0.25E-06 &  0.108009 & 14.98 & 11.76 &  9.53 &  8.19 &  7.52 &  574.2 &  566.0 &  464.3 &  367.7 &  106.66 &   18.76 &   4.778 \\  
0.50E-06 &  0.208266 & 16.02 & 12.47 &  9.96 &  8.46 &  7.61 &  747.2 &  786.7 &  698.1 &  592.6 &  173.15 &   31.57 &   8.023 \\  
0.10E-05 &  0.393236 & 17.91 & 13.78 & 10.77 &  8.99 &  7.85 &  913.5 & 1054.0 & 1018.6 &  923.7 &  281.23 &   52.72 &  13.390 \\  
0.20E-05 &  0.725089 & 21.26 & 16.12 & 12.23 &  9.97 &  8.40 &  955.8 & 1270.3 & 1375.0 & 1355.7 &  455.47 &   87.87 &  22.354 \\  
0.40E-05 &  1.306390 & 26.92 & 20.19 & 14.77 & 11.71 &  9.47 &  775.9 & 1287.4 & 1647.3 & 1839.5 &  746.08 &  149.40 &  38.197 \\  
0.60E-05 &  1.825759 & 31.81 & 23.78 & 16.99 & 13.21 & 10.42 &  587.2 & 1164.4 & 1699.8 & 2097.6 & 1002.97 &  206.71 &  53.101 \\  
0.10E-04 &  2.756648 & 40.25 & 30.11 & 20.72 & 15.67 & 12.01 &  341.1 &  893.9 & 1602.9 & 2320.3 & 1458.59 &  315.02 &  81.643 \\  

\hline
\end{tabular}
\label{tab1}
\end{table*}

\subsection{Caveats}

My expectation is that these tracks might be useful to identify AGB
stars in colour-colour and colour-magnitude diagrams that will become
available when SPITZER and ASTRO-F results on external galaxies become
available. Such a comparison might also be useful to provide
indications of the chemical type and dust optical depth. However, when
doing such comparisons some limitations of the models must be kept in mind.
\begin{itemize}

\item
AGB stars are variable, usually of the Semi-Regular or Mira type.
Mira variables can have pulsation amplitudes corresponding to 6-8 
magnitudes in $V$, and up to 2.5 mag in $K$, 1.5 mag in $M$ and
1.3 mag in $N$ (e.g. Le Bertre 1992, 1993, Groenewegen et al. 1998).
Little is known about the level of variability in the mid-IR, but for
example CW Leo (IRC +10~216), with a full-amplitude in $K$ of 2.0 mag
(Le Bertre 1992), still has an full-amplitude at 850 $\mu$m of 0.24 mag (Jenness et al. 2002).

\item
Other dust components may be present that are not taken into account
(MgS near 35 $\mu$m in carbon stars, e.g. Hony et al 2002, several
crystalline silicate complexes near 23, 28, 33, 40 and 60 $\mu$m, see
e.g. Molster et al. 2002) and that may influence some fluxes,
especially below $\sim 60 \mu m$.

\item
The dust shell may be non-spherical and scattered light might play a role.

\item
The models have been calculated assuming a constant mass loss rate. 
This may affect the far-IR colours (beyond $\sim 100\mu m$) either way. 

\item
The photospheric models used are appropriate for solar metallicities. 
This implies possible shifts in colour for low optical depths where
the photosphere dominate the colours in systems of non-solar metallicity.

\item
The post-AGB model have been calculated under the assumption that the
effective temperature and luminosity do not change over the time for
the dust shell to drift away.  For example, in the case of the 2650 K
effective temperature C-rich central star, with a luminosity of
3000 \lsol\ and 10 \ks\ expansion velocity it takes about 1100 year for
the dust shell to expand to an inner dust radius temperature of 100 K,
8300 yr to 50 K, and 600~000 yr to 10 K.

The transition timescale between the end of the AGB and the start of
the PN phase (typically assumed to start at 10~000 K) is highly
uncertain, and depends on the core mass (i.e. initial mass) of the
star (the larger, the faster the evolution). Typical values are
thought to be between 500 and 1500 years (e.g. Marigo et al. 2001, 2004).

This implies that only for inner dust radius temperatures above \more
150 K the flow times scale is short enough for the assumption of
constant luminosity and effective temperature likely to be valid.

\item
The final calibration may differ from the adopted one once the total throughput
of the system has been established in-orbit.

\end{itemize}

\section{Discussion}

Dust radiative transfer models for (post-)AGB stars are presented from
the optical to the far-infrared. The models are calculated in view of
the (upcoming) results from SPITZER and (hopefully) ASTRO-F. They may
be useful to identify AGB stars, and provide rough estimates of
spectral type, luminosity and mass loss rate. I warn again, though,
that once candidate AGB stars have been selected, the {\it best}
estimates for mass loss rate, luminosity and dust composition can only
be obtained by detailed fitting of the entire SED (and spectra if
available) of individual stars.

\begin{figure}
\includegraphics[width=85mm]{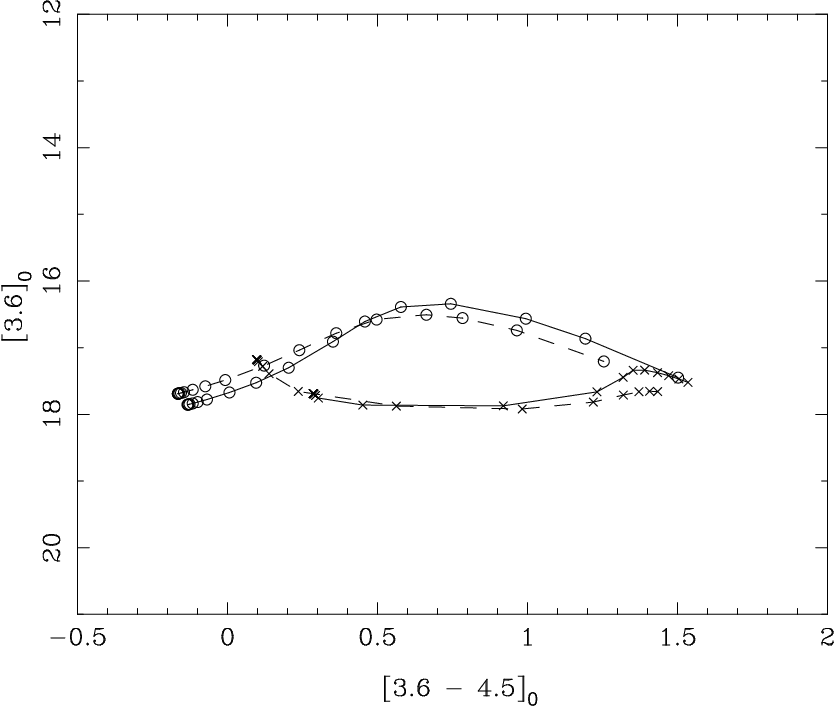}
\caption[]{
IRAC colour-magnitude diagram for models with 3000 \lsol\ at 932 Kpc. 
The following sequences of increasing mass loss are shown:
Carbon-rich AGB star with $T_{\rm eff}$ = 3600 K, and 85\% AMC + 15\% SiC (circles \& solid line),
Carbon-rich post-AGB star with $T_{\rm eff}$ = 2650 K, and 100\% AMC (crosses \& solid line),
Oxygen-rich AGB star with $T_{\rm eff}$ = 3297 K, and 60\% silicate + 40\% AlOx (circles \& dashed line), and
Oxygen-rich post-AGB star with $T_{\rm eff}$ = 2500 K, and 100\% silicate  (crosses \& dashed line).
}
\label{Fig-WLM}
\end{figure}

\begin{figure}
\includegraphics[width=85mm]{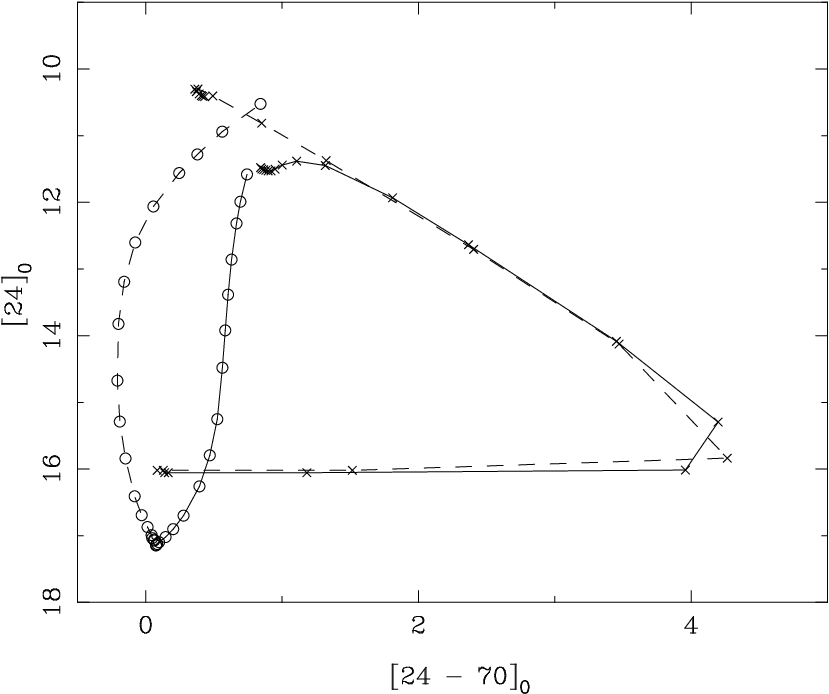}
\caption[]{
MIPS colour-magnitude diagram for models with 3000 \lsol\ at 785 Kpc. 
Model tracks as in Fig.~\ref{Fig-WLM}.
}
\label{Fig-M31}
\end{figure}

As an illustration, Figs.~\ref{Fig-WLM} and \ref{Fig-M31} show some
tracks in a colour-magnitude diagram for, respectively, SPITZER IRAC
filters for an AGB star of 3000 \lsol\ at a distance of 932 Kpc
(appropriate for WLM, McConnachie et al. 2005), and SPITZER MIPS
filters for an AGB star of 3000 \lsol\ at a distance of 785 Kpc
(appropriate for M31, McConnachie et al. 2005).
WLM and M31 are actual targets for those instruments according
to the Spitzer Reserved Observations Catalog. 
One can observe that the tracks of the C- and O-rich mass-loss
sequence largely overlap in [3.6-4.5] colour. Also the post-AGB tracks
cover the same range in this colour. On the other hand, the [3.6]
magnitude is a good indicator of the luminosity and the [3.6-4.5]
colours of the mass loss rate. By comparison, in Figure~\ref{Fig-M31}
C- and O-rich models separate, as well as AGB from post-AGB
models. This illustrates the usefulness of a 70 $\mu$m filter to
trace post-AGB evolution. The MIPS 24 $\mu$m filter traces the wing of
the silicate 18 $\mu$m feature and therefore the 24 $\mu$m magnitude
brightens quickly with the onset of mass-loss, so that the [24-70]
colour initially becomes negative for O-rich stars. On the other hand,
luminosity and mass loss rate are difficult to discriminate as the AGB
sequence is almost vertical in [24-70] colour.

\acknowledgements{
I would like to thanks Issei Yamamura (Institute of Space and
Astronomical Science, ISAS, Kanagawa, Japan) for pointing out the
availability of the ASTRO-F filter curves, and Leen Decin
(K.U. Leuven) for providing a MARCS code model atmosphere for Vega.
Joris Blommaert is thanked for reading an earlier version of the
paper.
} 
\\

{}

\appendix

\section{Tabular material}
\label{AppA}




\begin{table*}[H]
\caption{Optical, NIR, IRAC and MIPS magnitudes and fluxes for a C-rich AGB star, $T_{\rm eff}$ = 2650 K, 85\% AMC + 15\% SiC.}
\begin{tabular}{rrrrrrrrrrrrrr} \hline
\small

\mdot\   & $\tau$    &  $V$  &  $I$  &  $J$  &  $H$  &  $K$  &    3.6 &    4.5 &    5.8 &    8.0 &   24    &   70    & 160  \\ \hline

0.10E-09 &  0.000045 & 13.85 & 10.99 &  9.07 &  7.92 &  7.49 &  282.3 &  236.6 &  140.1 &   69.2 &   23.36 &    2.91 &   0.773 \\  
0.25E-09 &  0.000113 & 13.85 & 10.99 &  9.07 &  7.92 &  7.49 &  282.5 &  236.9 &  140.3 &   69.4 &   23.42 &    2.92 &   0.776 \\  
0.50E-09 &  0.000226 & 13.85 & 10.99 &  9.07 &  7.92 &  7.49 &  282.9 &  237.3 &  140.7 &   69.7 &   23.51 &    2.93 &   0.781 \\  
0.10E-08 &  0.000452 & 13.85 & 11.00 &  9.07 &  7.92 &  7.49 &  283.6 &  238.1 &  141.5 &   70.5 &   23.71 &    2.97 &   0.790 \\  
0.25E-08 &  0.001130 & 13.86 & 11.00 &  9.07 &  7.93 &  7.49 &  285.9 &  240.5 &  143.8 &   72.6 &   24.28 &    3.08 &   0.817 \\  
0.50E-08 &  0.002259 & 13.87 & 11.01 &  9.08 &  7.93 &  7.49 &  289.7 &  244.6 &  147.7 &   76.1 &   25.24 &    3.26 &   0.863 \\  
0.10E-07 &  0.004513 & 13.89 & 11.02 &  9.08 &  7.93 &  7.49 &  297.2 &  252.7 &  155.4 &   83.1 &   27.14 &    3.62 &   0.953 \\
\hline
\end{tabular}
\label{tab1}
\end{table*}

\begin{table*}
\caption{Optical, NIR, IRAC and MIPS magnitudes and fluxes for a C-rich AGB star, $T_{\rm eff}$ = 3600 K, 85\% AMC + 15\% SiC.}
\begin{tabular}{rrrrrrrrrrrrrr} \hline
\small

\mdot\   & $\tau$    &  $V$  &  $I$  &  $J$  &  $H$  &  $K$  &    3.6 &    4.5 &    5.8 &    8.0 &   24    &   70    & 160  \\ \hline

0.10E-09 &  0.000040 & 12.01 & 10.41 &  9.08 &  8.21 &  7.97 &  241.7 &  137.7 &   94.5 &   61.8 &    8.53 &    1.00 &   0.264 \\  
0.25E-09 &  0.000099 & 12.01 & 10.41 &  9.08 &  8.21 &  7.97 &  241.9 &  138.0 &   94.8 &   62.0 &    8.60 &    1.01 &   0.267 \\  
0.50E-09 &  0.000198 & 12.01 & 10.41 &  9.08 &  8.21 &  7.97 &  242.4 &  138.5 &   95.3 &   62.4 &    8.71 &    1.03 &   0.272 \\  
0.10E-08 &  0.000395 & 12.01 & 10.41 &  9.08 &  8.21 &  7.97 &  243.3 &  139.5 &   96.2 &   63.3 &    8.93 &    1.07 &   0.283 \\  
0.25E-08 &  0.000988 & 12.02 & 10.41 &  9.08 &  8.21 &  7.97 &  246.0 &  142.4 &   98.9 &   65.7 &    9.59 &    1.20 &   0.315 \\  
0.50E-08 &  0.001975 & 12.03 & 10.42 &  9.09 &  8.22 &  7.96 &  250.6 &  147.3 &  103.5 &   69.8 &   10.70 &    1.41 &   0.367 \\  
0.10E-07 &  0.003946 & 12.05 & 10.44 &  9.10 &  8.22 &  7.96 &  259.6 &  157.0 &  112.5 &   77.8 &   12.88 &    1.82 &   0.471 \\
\hline
\end{tabular}
\label{tab1}
\end{table*}

\begin{table*}
\caption{Optical, NIR, IRAC and MIPS magnitudes and fluxes for a C-rich AGB star, $T_{\rm eff}$ = 2650 K, 100\% AMC.}
\begin{tabular}{rrrrrrrrrrrrrr} \hline
\small

\mdot\   & $\tau$    &  $V$  &  $I$  &  $J$  &  $H$  &  $K$  &    3.6 &    4.5 &    5.8 &    8.0 &   24    &   70    & 160  \\ \hline

0.10E-09 &  0.000017 & 13.85 & 10.99 &  9.07 &  7.92 &  7.49 &  282.3 &  236.6 &  140.1 &   69.2 &   23.37 &    2.91 &   0.775 \\  
0.25E-09 &  0.000043 & 13.85 & 10.99 &  9.07 &  7.92 &  7.49 &  282.5 &  236.9 &  140.4 &   69.5 &   23.43 &    2.92 &   0.779 \\  
0.50E-09 &  0.000086 & 13.85 & 10.99 &  9.07 &  7.92 &  7.49 &  282.9 &  237.4 &  140.8 &   69.9 &   23.54 &    2.94 &   0.787 \\  
0.10E-08 &  0.000172 & 13.85 & 11.00 &  9.07 &  7.92 &  7.49 &  283.7 &  238.2 &  141.7 &   70.7 &   23.75 &    2.99 &   0.802 \\  
0.25E-08 &  0.000429 & 13.86 & 11.00 &  9.07 &  7.93 &  7.49 &  286.2 &  240.9 &  144.4 &   73.2 &   24.40 &    3.13 &   0.848 \\  
0.50E-08 &  0.000857 & 13.87 & 11.01 &  9.08 &  7.93 &  7.49 &  290.2 &  245.4 &  148.8 &   77.3 &   25.48 &    3.35 &   0.923 \\  
0.10E-07 &  0.001712 & 13.90 & 11.03 &  9.09 &  7.93 &  7.48 &  298.2 &  254.2 &  157.6 &   85.5 &   27.61 &    3.80 &   1.075 \\
\hline
\end{tabular}
\label{tab1}
\end{table*}

\begin{table*}
\caption{Optical, NIR, IRAC and MIPS magnitudes and fluxes for a C-rich AGB star, $T_{\rm eff}$ = 3600 K, 100\% AMC.}
\begin{tabular}{rrrrrrrrrrrrrr} \hline
\small

\mdot\   & $\tau$    &  $V$  &  $I$  &  $J$  &  $H$  &  $K$  &    3.6 &    4.5 &    5.8 &    8.0 &   24    &   70    & 160  \\ \hline

0.10E-09 &  0.000015 & 12.01 & 10.41 &  9.08 &  8.21 &  7.97 &  241.7 &  137.8 &   94.6 &   61.8 &    8.54 &    1.00 &   0.266 \\  
0.25E-09 &  0.000037 & 12.01 & 10.41 &  9.08 &  8.21 &  7.97 &  242.0 &  138.1 &   94.9 &   62.1 &    8.61 &    1.02 &   0.271 \\  
0.50E-09 &  0.000074 & 12.01 & 10.41 &  9.08 &  8.21 &  7.97 &  242.5 &  138.6 &   95.4 &   62.6 &    8.74 &    1.04 &   0.280 \\  
0.10E-08 &  0.000149 & 12.01 & 10.41 &  9.08 &  8.21 &  7.97 &  243.4 &  139.7 &   96.5 &   63.6 &    8.99 &    1.10 &   0.297 \\  
0.25E-08 &  0.000372 & 12.02 & 10.41 &  9.08 &  8.21 &  7.97 &  246.4 &  142.9 &   99.6 &   66.4 &    9.74 &    1.25 &   0.350 \\  
0.50E-08 &  0.000743 & 12.03 & 10.42 &  9.09 &  8.22 &  7.96 &  251.3 &  148.3 &  104.8 &   71.3 &   10.99 &    1.52 &   0.438 \\  
0.10E-07 &  0.001485 & 12.05 & 10.44 &  9.10 &  8.22 &  7.96 &  260.9 &  159.0 &  115.1 &   80.8 &   13.46 &    2.04 &   0.613 \\
\hline
\end{tabular}
\label{tab1}
\end{table*}

\begin{table*}
\caption{Optical, NIR, IRAC and MIPS magnitudes and fluxes for a O-rich AGB star, $T_{\rm eff}$ = 3850 K, 100\% AlOx.}
\begin{tabular}{rrrrrrrrrrrrrr} \hline
\small

\mdot\   & $\tau$    &  $V$  &  $I$  &  $J$  &  $H$  &  $K$  &    3.6 &    4.5 &    5.8 &    8.0 &   24    &   70    & 160  \\ \hline

0.10E-09 &  0.000200 & 11.84 & 10.19 &  9.08 &  8.21 &  8.08 &  177.3 &   98.3 &   69.8 &   42.2 &    5.35 &    0.61 &   0.150 \\  
0.25E-09 &  0.000499 & 11.84 & 10.19 &  9.08 &  8.21 &  8.08 &  177.4 &   98.3 &   69.8 &   42.3 &    5.45 &    0.62 &   0.153 \\  
0.50E-09 &  0.000998 & 11.84 & 10.19 &  9.08 &  8.21 &  8.08 &  177.4 &   98.4 &   69.8 &   42.4 &    5.61 &    0.64 &   0.158 \\  
0.10E-08 &  0.001996 & 11.84 & 10.19 &  9.08 &  8.21 &  8.08 &  177.5 &   98.5 &   70.0 &   42.6 &    5.93 &    0.68 &   0.167 \\  
0.25E-08 &  0.004989 & 11.84 & 10.19 &  9.09 &  8.21 &  8.08 &  177.7 &   98.7 &   70.3 &   43.4 &    6.90 &    0.78 &   0.196 \\  
0.50E-08 &  0.009974 & 11.85 & 10.19 &  9.09 &  8.21 &  8.08 &  178.1 &   99.1 &   70.8 &   44.7 &    8.52 &    0.95 &   0.244 \\  
0.10E-07 &  0.019932 & 11.85 & 10.20 &  9.09 &  8.21 &  8.08 &  178.8 &  100.0 &   71.9 &   47.3 &   11.74 &    1.30 &   0.339 \\
\hline
\end{tabular}
\label{tab1}
\end{table*}

\begin{table*}
\caption{Optical, NIR, IRAC and MIPS magnitudes and fluxes for a O-rich AGB star, $T_{\rm eff}$ = 3297 K, 100\% AlOx.}
\begin{tabular}{rrrrrrrrrrrrrr} \hline
\small

\mdot\   & $\tau$    &  $V$  &  $I$  &  $J$  &  $H$  &  $K$  &    3.6 &    4.5 &    5.8 &    8.0 &   24    &   70    & 160  \\ \hline

0.10E-09 &  0.000242 & 14.31 & 11.11 &  8.88 &  7.85 &  7.62 &  281.1 &  155.5 &  112.1 &   68.8 &    9.11 &    1.05 &   0.258 \\  
0.25E-09 &  0.000604 & 14.31 & 11.11 &  8.88 &  7.85 &  7.62 &  281.1 &  155.5 &  112.1 &   68.9 &    9.19 &    1.06 &   0.261 \\  
0.50E-09 &  0.001209 & 14.31 & 11.11 &  8.88 &  7.85 &  7.62 &  281.2 &  155.5 &  112.1 &   69.0 &    9.32 &    1.08 &   0.265 \\  
0.10E-08 &  0.002417 & 14.31 & 11.11 &  8.88 &  7.85 &  7.62 &  281.2 &  155.6 &  112.2 &   69.2 &    9.58 &    1.10 &   0.272 \\  
0.25E-08 &  0.006041 & 14.31 & 11.11 &  8.88 &  7.85 &  7.62 &  281.4 &  155.8 &  112.5 &   69.8 &   10.37 &    1.19 &   0.296 \\  
0.50E-08 &  0.012076 & 14.32 & 11.11 &  8.88 &  7.85 &  7.62 &  281.6 &  156.1 &  112.9 &   70.8 &   11.68 &    1.33 &   0.335 \\  
0.10E-07 &  0.024129 & 14.33 & 11.12 &  8.88 &  7.85 &  7.62 &  282.2 &  156.7 &  113.8 &   72.9 &   14.30 &    1.61 &   0.413 \\
\hline
\end{tabular}
\label{tab1}
\end{table*}

\begin{table*}
\caption{Optical, NIR, IRAC and MIPS magnitudes and fluxes for a O-rich AGB star, $T_{\rm eff}$ = 2500 K, 100\% AlOx.}
\begin{tabular}{rrrrrrrrrrrrrr} \hline
\small

\mdot\   & $\tau$    &  $V$  &  $I$  &  $J$  &  $H$  &  $K$  &    3.6 &    4.5 &    5.8 &    8.0 &   24    &   70    & 160  \\ \hline

0.10E-09 &  0.000258 & 17.04 & 12.72 &  8.67 &  7.87 &  7.44 &  451.6 &  318.0 &  247.5 &  152.9 &   24.16 &    2.86 &   0.703 \\  
0.25E-09 &  0.000644 & 17.04 & 12.72 &  8.67 &  7.87 &  7.44 &  451.6 &  318.1 &  247.5 &  153.0 &   24.23 &    2.86 &   0.705 \\  
0.50E-09 &  0.001287 & 17.04 & 12.72 &  8.67 &  7.87 &  7.44 &  451.6 &  318.1 &  247.5 &  153.0 &   24.34 &    2.88 &   0.709 \\  
0.10E-08 &  0.002575 & 17.04 & 12.72 &  8.67 &  7.87 &  7.44 &  451.7 &  318.1 &  247.6 &  153.2 &   24.57 &    2.90 &   0.716 \\  
0.25E-08 &  0.006435 & 17.04 & 12.72 &  8.67 &  7.87 &  7.44 &  451.8 &  318.3 &  247.8 &  153.7 &   25.25 &    2.98 &   0.737 \\  
0.50E-08 &  0.012865 & 17.05 & 12.72 &  8.68 &  7.87 &  7.44 &  452.0 &  318.5 &  248.1 &  154.5 &   26.40 &    3.10 &   0.773 \\  
0.10E-07 &  0.025708 & 17.06 & 12.73 &  8.68 &  7.87 &  7.44 &  452.3 &  319.0 &  248.8 &  156.2 &   28.68 &    3.36 &   0.844 \\
\hline
\end{tabular}
\label{tab1}
\end{table*}

\begin{table*}
\caption{Optical, NIR, IRAC and MIPS magnitudes and fluxes for a O-rich AGB star, $T_{\rm eff}$ = 3850 K, 60\% Silicate + 40\% Alox.}
\begin{tabular}{rrrrrrrrrrrrrr} \hline
\small

\mdot\   & $\tau$    &  $V$  &  $I$  &  $J$  &  $H$  &  $K$  &    3.6 &    4.5 &    5.8 &    8.0 &   24    &   70    & 160  \\ \hline

0.10E-09 &  0.000145 & 11.84 & 10.19 &  9.08 &  8.21 &  8.08 &  177.4 &   98.4 &   69.8 &   42.2 &    5.38 &    0.61 &   0.150 \\  
0.25E-09 &  0.000362 & 11.84 & 10.19 &  9.08 &  8.21 &  8.08 &  177.5 &   98.5 &   69.9 &   42.4 &    5.53 &    0.63 &   0.154 \\  
0.50E-09 &  0.000725 & 11.84 & 10.19 &  9.08 &  8.21 &  8.08 &  177.8 &   98.7 &   70.1 &   42.7 &    5.77 &    0.65 &   0.160 \\  
0.10E-08 &  0.001449 & 11.84 & 10.19 &  9.09 &  8.21 &  8.08 &  178.2 &   99.2 &   70.5 &   43.3 &    6.26 &    0.69 &   0.172 \\  
0.25E-08 &  0.003621 & 11.85 & 10.19 &  9.09 &  8.21 &  8.08 &  179.5 &  100.4 &   71.7 &   45.0 &    7.71 &    0.81 &   0.207 \\  
0.50E-08 &  0.007239 & 11.86 & 10.20 &  9.09 &  8.21 &  8.07 &  181.6 &  102.5 &   73.7 &   47.9 &   10.11 &    1.02 &   0.266 \\  
0.10E-07 &  0.014465 & 11.88 & 10.21 &  9.09 &  8.21 &  8.07 &  185.9 &  106.8 &   77.7 &   53.7 &   14.89 &    1.42 &   0.384 \\
\hline
\end{tabular}
\label{tab1}
\end{table*}

\begin{table*}
\caption{Optical, NIR, IRAC and MIPS magnitudes and fluxes for a O-rich AGB star, $T_{\rm eff}$ = 3297 K, 60\% Silicate + 40\% Alox.}
\begin{tabular}{rrrrrrrrrrrrrr} \hline
\small

\mdot\   & $\tau$    &  $V$  &  $I$  &  $J$  &  $H$  &  $K$  &    3.6 &    4.5 &    5.8 &    8.0 &   24    &   70    & 160  \\ \hline

0.10E-09 &  0.000177 & 14.31 & 11.11 &  8.88 &  7.85 &  7.62 &  281.2 &  155.5 &  112.1 &   68.9 &    9.14 &    1.06 &   0.259 \\  
0.25E-09 &  0.000443 & 14.31 & 11.11 &  8.88 &  7.85 &  7.62 &  281.3 &  155.6 &  112.2 &   69.0 &    9.25 &    1.07 &   0.262 \\  
0.50E-09 &  0.000886 & 14.31 & 11.11 &  8.88 &  7.85 &  7.62 &  281.4 &  155.8 &  112.4 &   69.3 &    9.45 &    1.08 &   0.266 \\  
0.10E-08 &  0.001772 & 14.31 & 11.11 &  8.88 &  7.85 &  7.62 &  281.7 &  156.1 &  112.7 &   69.7 &    9.84 &    1.12 &   0.276 \\  
0.25E-08 &  0.004427 & 14.32 & 11.11 &  8.88 &  7.85 &  7.62 &  282.7 &  157.1 &  113.6 &   71.1 &   11.01 &    1.22 &   0.305 \\  
0.50E-08 &  0.008850 & 14.33 & 11.12 &  8.88 &  7.85 &  7.62 &  284.2 &  158.7 &  115.2 &   73.4 &   12.96 &    1.38 &   0.353 \\  
0.10E-07 &  0.017680 & 14.36 & 11.13 &  8.89 &  7.86 &  7.62 &  287.4 &  162.0 &  118.3 &   78.0 &   16.84 &    1.71 &   0.449 \\
\hline
\end{tabular}
\label{tab1}
\end{table*}

\begin{table*}
\caption{Optical, NIR, IRAC and MIPS magnitudes and fluxes for a O-rich AGB star, $T_{\rm eff}$ = 2500 K, 60\% Silicate + 40\% Alox.}
\begin{tabular}{rrrrrrrrrrrrrr} \hline
\small

\mdot\   & $\tau$    &  $V$  &  $I$  &  $J$  &  $H$  &  $K$  &    3.6 &    4.5 &    5.8 &    8.0 &   24    &   70    & 160  \\ \hline

0.10E-09 &  0.000189 & 17.04 & 12.72 &  8.67 &  7.87 &  7.44 &  451.6 &  318.1 &  247.5 &  153.0 &   24.18 &    2.86 &   0.704 \\  
0.25E-09 &  0.000473 & 17.04 & 12.72 &  8.67 &  7.87 &  7.44 &  451.7 &  318.2 &  247.6 &  153.1 &   24.29 &    2.87 &   0.706 \\  
0.50E-09 &  0.000945 & 17.04 & 12.72 &  8.67 &  7.87 &  7.44 &  451.8 &  318.3 &  247.7 &  153.3 &   24.46 &    2.88 &   0.711 \\  
0.10E-08 &  0.001890 & 17.04 & 12.72 &  8.67 &  7.87 &  7.44 &  452.0 &  318.5 &  248.0 &  153.7 &   24.80 &    2.91 &   0.719 \\  
0.25E-08 &  0.004724 & 17.05 & 12.73 &  8.68 &  7.87 &  7.44 &  452.7 &  319.3 &  248.7 &  154.8 &   25.84 &    3.00 &   0.746 \\  
0.50E-08 &  0.009443 & 17.07 & 12.73 &  8.68 &  7.87 &  7.44 &  453.9 &  320.6 &  250.0 &  156.8 &   27.57 &    3.15 &   0.789 \\  
0.10E-07 &  0.018867 & 17.09 & 12.74 &  8.69 &  7.88 &  7.44 &  456.1 &  323.1 &  252.5 &  160.7 &   31.01 &    3.45 &   0.877 \\
\hline
\end{tabular}
\label{tab1}
\end{table*}

\begin{table*}
\caption{Optical, NIR, IRAC and MIPS magnitudes and fluxes for a O-rich AGB star, $T_{\rm eff}$ = 3850 K, 100\% Silicate.}
\begin{tabular}{rrrrrrrrrrrrrr} \hline
\small

\mdot\   & $\tau$    &  $V$  &  $I$  &  $J$  &  $H$  &  $K$  &    3.6 &    4.5 &    5.8 &    8.0 &   24    &   70    & 160  \\ \hline

0.10E-09 &  0.000113 & 11.84 & 10.19 &  9.08 &  8.21 &  8.08 &  177.4 &   98.4 &   69.8 &   42.2 &    5.37 &    0.61 &   0.150 \\  
0.25E-09 &  0.000283 & 11.84 & 10.19 &  9.08 &  8.21 &  8.08 &  177.4 &   98.4 &   69.9 &   42.4 &    5.50 &    0.62 &   0.153 \\  
0.50E-09 &  0.000566 & 11.84 & 10.19 &  9.08 &  8.21 &  8.08 &  177.5 &   98.6 &   70.0 &   42.6 &    5.71 &    0.64 &   0.158 \\  
0.10E-08 &  0.001132 & 11.84 & 10.19 &  9.09 &  8.21 &  8.08 &  177.7 &   98.8 &   70.3 &   43.1 &    6.14 &    0.67 &   0.167 \\  
0.25E-08 &  0.002830 & 11.84 & 10.19 &  9.09 &  8.21 &  8.08 &  178.3 &   99.6 &   71.1 &   44.5 &    7.42 &    0.76 &   0.196 \\  
0.50E-08 &  0.005659 & 11.85 & 10.19 &  9.09 &  8.21 &  8.08 &  179.3 &  100.9 &   72.5 &   46.8 &    9.55 &    0.92 &   0.245 \\  
0.10E-07 &  0.011312 & 11.86 & 10.20 &  9.09 &  8.21 &  8.08 &  181.3 &  103.5 &   75.4 &   51.4 &   13.80 &    1.23 &   0.341 \\
\hline
\end{tabular}
\label{tab1}
\end{table*}

\begin{table*}
\caption{Optical, NIR, IRAC and MIPS magnitudes and fluxes for a O-rich AGB star, $T_{\rm eff}$ = 3297 K, 100\% Silicate.}
\begin{tabular}{rrrrrrrrrrrrrr} \hline
\small

\mdot\   & $\tau$    &  $V$  &  $I$  &  $J$  &  $H$  &  $K$  &    3.6 &    4.5 &    5.8 &    8.0 &   24    &   70    & 160  \\ \hline

0.10E-09 &  0.000134 & 14.31 & 11.11 &  8.88 &  7.85 &  7.62 &  281.1 &  155.5 &  112.1 &   68.9 &    9.13 &    1.05 &   0.258 \\  
0.25E-09 &  0.000334 & 14.31 & 11.11 &  8.88 &  7.85 &  7.62 &  281.2 &  155.5 &  112.2 &   69.0 &    9.24 &    1.06 &   0.261 \\  
0.50E-09 &  0.000668 & 14.31 & 11.11 &  8.88 &  7.85 &  7.62 &  281.3 &  155.7 &  112.3 &   69.2 &    9.42 &    1.07 &   0.265 \\  
0.10E-08 &  0.001337 & 14.31 & 11.11 &  8.88 &  7.85 &  7.62 &  281.4 &  155.9 &  112.5 &   69.6 &    9.78 &    1.10 &   0.273 \\  
0.25E-08 &  0.003342 & 14.31 & 11.11 &  8.88 &  7.85 &  7.62 &  281.8 &  156.5 &  113.2 &   70.7 &   10.86 &    1.18 &   0.298 \\  
0.50E-08 &  0.006681 & 14.32 & 11.11 &  8.88 &  7.85 &  7.62 &  282.6 &  157.5 &  114.4 &   72.7 &   12.66 &    1.31 &   0.338 \\  
0.10E-07 &  0.013355 & 14.33 & 11.12 &  8.89 &  7.86 &  7.62 &  284.1 &  159.6 &  116.7 &   76.6 &   16.25 &    1.58 &   0.420 \\
\hline
\end{tabular}
\label{tab1}
\end{table*}

\begin{table*}
\caption{Optical, NIR, IRAC and MIPS magnitudes and fluxes for a O-rich AGB star, $T_{\rm eff}$ = 2500 K, 100\% Silicate.}
\begin{tabular}{rrrrrrrrrrrrrr} \hline
\small

\mdot\   & $\tau$    &  $V$  &  $I$  &  $J$  &  $H$  &  $K$  &    3.6 &    4.5 &    5.8 &    8.0 &   24    &   70    & 160  \\ \hline

0.10E-09 &  0.000144 & 17.04 & 12.72 &  8.67 &  7.87 &  7.44 &  451.6 &  318.1 &  247.5 &  153.0 &   24.18 &    2.85 &   0.703 \\  
0.25E-09 &  0.000361 & 17.04 & 12.72 &  8.67 &  7.87 &  7.44 &  451.6 &  318.1 &  247.6 &  153.1 &   24.28 &    2.86 &   0.706 \\  
0.50E-09 &  0.000722 & 17.04 & 12.72 &  8.67 &  7.87 &  7.44 &  451.7 &  318.2 &  247.7 &  153.2 &   24.44 &    2.87 &   0.709 \\  
0.10E-08 &  0.001445 & 17.04 & 12.72 &  8.67 &  7.87 &  7.44 &  451.8 &  318.4 &  247.9 &  153.6 &   24.77 &    2.90 &   0.717 \\  
0.25E-08 &  0.003611 & 17.05 & 12.72 &  8.67 &  7.87 &  7.44 &  452.1 &  318.9 &  248.5 &  154.6 &   25.76 &    2.97 &   0.739 \\  
0.50E-08 &  0.007219 & 17.05 & 12.73 &  8.68 &  7.87 &  7.44 &  452.6 &  319.7 &  249.5 &  156.4 &   27.40 &    3.09 &   0.777 \\  
0.10E-07 &  0.014430 & 17.06 & 12.73 &  8.68 &  7.88 &  7.44 &  453.7 &  321.4 &  251.5 &  159.8 &   30.69 &    3.34 &   0.852 \\
\hline
\end{tabular}
\label{tab1}
\end{table*}


\begin{table*}
\caption{ASTRO-F fluxes for a C-rich AGB star, $T_{\rm eff}$ = 2650 K, 85\% AMC + 15\% SiC.}
\setlength{\tabcolsep}{1.15mm}
\begin{tabular}{rrrrrrrrrrrrrrr} \hline

\mdot\   & $\tau$    & IRC\_N2 & IRC\_N3 & IRC\_N4 & S7  &   S9W  &   S11  &  L15   &  L18W  &  L24   &  N60  & WIDE-S  & WIDE-L  & N160  \\ \hline

0.10E-09 &  0.000045 &  637.4 &  296.2 &  255.2 &   90.4 &   65.8 &   39.5 &   25.3 &   21.1 &   23.7 &    3.40 &    2.26 &    0.83 &    0.66 \\  
0.25E-09 &  0.000113 &  637.5 &  296.4 &  255.5 &   90.6 &   66.0 &   39.7 &   25.4 &   21.1 &   23.7 &    3.41 &    2.27 &    0.83 &    0.67 \\  
0.50E-09 &  0.000226 &  637.6 &  296.7 &  255.9 &   91.0 &   66.4 &   40.1 &   25.5 &   21.3 &   23.8 &    3.43 &    2.28 &    0.84 &    0.67 \\  
0.10E-08 &  0.000452 &  637.8 &  297.3 &  256.7 &   91.8 &   67.3 &   40.8 &   25.9 &   21.5 &   24.0 &    3.47 &    2.31 &    0.85 &    0.68 \\  
0.25E-08 &  0.001130 &  638.5 &  299.2 &  259.2 &   94.3 &   69.9 &   42.9 &   27.0 &   22.2 &   24.6 &    3.60 &    2.40 &    0.88 &    0.70 \\  
0.50E-08 &  0.002259 &  639.7 &  302.4 &  263.4 &   98.3 &   74.1 &   46.3 &   28.8 &   23.4 &   25.6 &    3.80 &    2.54 &    0.93 &    0.74 \\  
0.10E-07 &  0.004513 &  642.0 &  308.6 &  271.7 &  106.4 &   82.6 &   53.2 &   32.3 &   25.8 &   27.5 &    4.21 &    2.83 &    1.03 &    0.81 \\
\hline
\end{tabular}
\label{tab1}
\end{table*}

\begin{table*}
\caption{ASTRO-F fluxes for a C-rich AGB star, $T_{\rm eff}$ = 3600 K, 85\% AMC + 15\% SiC.}
\setlength{\tabcolsep}{1.15mm}
\begin{tabular}{rrrrrrrrrrrrrrr} \hline
\mdot\   & $\tau$    & IRC\_N2 & IRC\_N3 & IRC\_N4 & S7  &   S9W  &   S11  &  L15   &  L18W  &  L24   &  N60  & WIDE-S  & WIDE-L  & N160  \\ \hline

0.10E-09 &  0.000040 &  404.9 &  262.9 &  160.5 &   78.4 &   57.1 &   33.6 &   19.7 &   12.5 &    8.7 &    1.17 &    0.77 &    0.28 &    0.23 \\  
0.25E-09 &  0.000099 &  405.0 &  263.1 &  160.8 &   78.7 &   57.4 &   33.9 &   19.9 &   12.6 &    8.8 &    1.18 &    0.78 &    0.29 &    0.23 \\  
0.50E-09 &  0.000198 &  405.3 &  263.5 &  161.3 &   79.2 &   57.9 &   34.3 &   20.1 &   12.7 &    8.9 &    1.21 &    0.80 &    0.29 &    0.23 \\  
0.10E-08 &  0.000395 &  405.7 &  264.3 &  162.3 &   80.1 &   58.9 &   35.1 &   20.5 &   13.0 &    9.1 &    1.25 &    0.84 &    0.30 &    0.24 \\  
0.25E-08 &  0.000988 &  407.1 &  266.6 &  165.3 &   82.9 &   61.8 &   37.5 &   21.7 &   13.8 &    9.8 &    1.40 &    0.94 &    0.34 &    0.27 \\  
0.50E-08 &  0.001975 &  409.4 &  270.4 &  170.3 &   87.6 &   66.8 &   41.5 &   23.8 &   15.2 &   10.9 &    1.64 &    1.10 &    0.40 &    0.31 \\  
0.10E-07 &  0.003946 &  413.9 &  278.0 &  180.2 &   96.9 &   76.6 &   49.5 &   27.9 &   17.9 &   13.2 &    2.11 &    1.44 &    0.51 &    0.40 \\
\hline
\end{tabular}
\label{tab1}
\end{table*}

\begin{table*}
\caption{ASTRO-F fluxes for a C-rich AGB star, $T_{\rm eff}$ = 2650 K, 100\% AMC.}
\setlength{\tabcolsep}{1.15mm}
\begin{tabular}{rrrrrrrrrrrrrrr} \hline
\mdot\   & $\tau$    & IRC\_N2 & IRC\_N3 & IRC\_N4 & S7  &   S9W  &   S11  &  L15   &  L18W  &  L24   &  N60  & WIDE-S  & WIDE-L  & N160  \\ \hline

0.10E-09 &  0.000017 &  637.4 &  296.2 &  255.2 &   90.4 &   65.8 &   39.5 &   25.3 &   21.1 &   23.7 &    3.40 &    2.26 &    0.83 &    0.66 \\  
0.25E-09 &  0.000043 &  637.5 &  296.4 &  255.5 &   90.7 &   66.0 &   39.6 &   25.4 &   21.2 &   23.7 &    3.41 &    2.27 &    0.83 &    0.67 \\  
0.50E-09 &  0.000086 &  637.6 &  296.7 &  256.0 &   91.2 &   66.4 &   39.9 &   25.6 &   21.3 &   23.8 &    3.44 &    2.29 &    0.84 &    0.67 \\  
0.10E-08 &  0.000172 &  637.9 &  297.4 &  256.9 &   92.1 &   67.2 &   40.5 &   25.9 &   21.6 &   24.1 &    3.49 &    2.33 &    0.86 &    0.69 \\  
0.25E-08 &  0.000429 &  638.6 &  299.4 &  259.6 &   95.0 &   69.7 &   42.1 &   27.1 &   22.3 &   24.7 &    3.64 &    2.44 &    0.91 &    0.73 \\  
0.50E-08 &  0.000857 &  639.9 &  302.7 &  264.2 &   99.7 &   73.8 &   44.9 &   28.9 &   23.6 &   25.8 &    3.90 &    2.64 &    0.99 &    0.79 \\  
0.10E-07 &  0.001712 &  642.3 &  309.3 &  273.3 &  109.1 &   81.9 &   50.4 &   32.7 &   26.2 &   28.0 &    4.40 &    3.02 &    1.15 &    0.92 \\
\hline
\end{tabular}
\label{tab1}
\end{table*}

\begin{table*}
\caption{ASTRO-F fluxes for a C-rich AGB star, $T_{\rm eff}$ = 3600 K, 100\% AMC.}
\setlength{\tabcolsep}{1.15mm}
\begin{tabular}{rrrrrrrrrrrrrrr} \hline
\mdot\   & $\tau$    & IRC\_N2 & IRC\_N3 & IRC\_N4 & S7  &   S9W  &   S11  &  L15   &  L18W  &  L24   &  N60  & WIDE-S  & WIDE-L  & N160  \\ \hline

0.10E-09 &  0.000015 &  404.9 &  262.9 &  160.5 &   78.5 &   57.1 &   33.6 &   19.8 &   12.5 &    8.7 &    1.17 &    0.78 &    0.28 &    0.23 \\  
0.25E-09 &  0.000037 &  405.0 &  263.2 &  160.8 &   78.8 &   57.4 &   33.8 &   19.9 &   12.6 &    8.8 &    1.19 &    0.79 &    0.29 &    0.23 \\  
0.50E-09 &  0.000074 &  405.3 &  263.6 &  161.4 &   79.3 &   57.8 &   34.1 &   20.1 &   12.7 &    9.0 &    1.22 &    0.81 &    0.30 &    0.24 \\  
0.10E-08 &  0.000149 &  405.8 &  264.4 &  162.5 &   80.5 &   58.8 &   34.8 &   20.5 &   13.0 &    9.2 &    1.28 &    0.86 &    0.32 &    0.26 \\  
0.25E-08 &  0.000372 &  407.3 &  266.9 &  165.8 &   83.8 &   61.7 &   36.7 &   21.9 &   14.0 &   10.0 &    1.45 &    0.99 &    0.38 &    0.30 \\  
0.50E-08 &  0.000743 &  409.7 &  271.0 &  171.3 &   89.3 &   66.4 &   39.9 &   24.0 &   15.5 &   11.2 &    1.75 &    1.21 &    0.47 &    0.38 \\  
0.10E-07 &  0.001485 &  414.5 &  279.0 &  182.2 &  100.2 &   75.8 &   46.3 &   28.4 &   18.5 &   13.8 &    2.33 &    1.65 &    0.66 &    0.53 \\
\hline
\end{tabular}
\label{tab1}
\end{table*}

\begin{table*}
\caption{ASTRO-F fluxes for a O-rich AGB star, $T_{\rm eff}$ = 3850 K, 100\% AlOx.}
\setlength{\tabcolsep}{1.15mm}
\begin{tabular}{rrrrrrrrrrrrrrr} \hline
\mdot\   & $\tau$    & IRC\_N2 & IRC\_N3 & IRC\_N4 & S7  &   S9W  &   S11  &  L15   &  L18W  &  L24   &  N60  & WIDE-S  & WIDE-L  & N160  \\ \hline

0.10E-09 &  0.000200 &  360.6 &  198.0 &  115.6 &   54.7 &   38.3 &   21.7 &   12.5 &    7.9 &    5.5 &    0.72 &    0.48 &    0.16 &    0.13 \\  
0.25E-09 &  0.000499 &  360.6 &  198.1 &  115.7 &   54.8 &   38.5 &   22.0 &   12.9 &    8.1 &    5.6 &    0.73 &    0.48 &    0.17 &    0.13 \\  
0.50E-09 &  0.000998 &  360.6 &  198.1 &  115.7 &   54.9 &   38.8 &   22.5 &   13.4 &    8.4 &    5.8 &    0.75 &    0.50 &    0.17 &    0.13 \\  
0.10E-08 &  0.001996 &  360.7 &  198.2 &  115.8 &   55.1 &   39.5 &   23.4 &   14.6 &    9.0 &    6.1 &    0.79 &    0.53 &    0.18 &    0.14 \\  
0.25E-08 &  0.004989 &  360.9 &  198.4 &  116.0 &   55.8 &   41.4 &   26.3 &   17.9 &   11.0 &    7.1 &    0.91 &    0.61 &    0.21 &    0.17 \\  
0.50E-08 &  0.009974 &  361.2 &  198.7 &  116.5 &   56.9 &   44.7 &   31.1 &   23.5 &   14.1 &    8.8 &    1.10 &    0.75 &    0.27 &    0.21 \\  
0.10E-07 &  0.019932 &  361.9 &  199.4 &  117.3 &   59.1 &   51.1 &   40.6 &   34.6 &   20.5 &   12.1 &    1.49 &    1.04 &    0.37 &    0.29 \\
\hline
\end{tabular}
\label{tab1}
\end{table*}

\clearpage

\begin{table*}
\caption{ASTRO-F fluxes for a O-rich AGB star, $T_{\rm eff}$ = 3297 K, 100\% AlOx.}
\setlength{\tabcolsep}{1.15mm}
\begin{tabular}{rrrrrrrrrrrrrrr} \hline
\mdot\   & $\tau$    & IRC\_N2 & IRC\_N3 & IRC\_N4 & S7  &   S9W  &   S11  &  L15   &  L18W  &  L24   &  N60  & WIDE-S  & WIDE-L  & N160  \\ \hline

0.10E-09 &  0.000242 &  525.1 &  294.4 &  185.0 &   89.3 &   63.1 &   36.1 &   21.1 &   13.4 &    9.3 &    1.23 &    0.82 &    0.28 &    0.22 \\  
0.25E-09 &  0.000604 &  525.1 &  294.4 &  185.1 &   89.3 &   63.3 &   36.3 &   21.4 &   13.5 &    9.4 &    1.24 &    0.83 &    0.28 &    0.22 \\  
0.50E-09 &  0.001209 &  525.1 &  294.4 &  185.1 &   89.4 &   63.5 &   36.7 &   21.8 &   13.8 &    9.6 &    1.26 &    0.84 &    0.29 &    0.23 \\  
0.10E-08 &  0.002417 &  525.1 &  294.5 &  185.2 &   89.6 &   64.1 &   37.5 &   22.7 &   14.3 &    9.8 &    1.29 &    0.86 &    0.30 &    0.23 \\  
0.25E-08 &  0.006041 &  525.2 &  294.6 &  185.4 &   90.1 &   65.6 &   39.8 &   25.5 &   15.8 &   10.6 &    1.39 &    0.93 &    0.32 &    0.25 \\  
0.50E-08 &  0.012076 &  525.3 &  294.8 &  185.7 &   91.0 &   68.2 &   43.6 &   30.0 &   18.4 &   12.0 &    1.55 &    1.04 &    0.37 &    0.29 \\  
0.10E-07 &  0.024129 &  525.6 &  295.3 &  186.3 &   92.7 &   73.4 &   51.2 &   38.9 &   23.5 &   14.7 &    1.87 &    1.28 &    0.45 &    0.35 \\
\hline
\end{tabular}
\label{tab1}
\end{table*}

\begin{table*}
\caption{ASTRO-F fluxes for a O-rich AGB star, $T_{\rm eff}$ = 2500 K, 100\% AlOx.}
\setlength{\tabcolsep}{1.15mm}
\begin{tabular}{rrrrrrrrrrrrrrr} \hline
\mdot\   & $\tau$    & IRC\_N2 & IRC\_N3 & IRC\_N4 & S7  &   S9W  &   S11  &  L15   &  L18W  &  L24   &  N60  & WIDE-S  & WIDE-L  & N160  \\ \hline

0.10E-09 &  0.000258 &  667.9 &  458.3 &  362.0 &  198.2 &  143.8 &   86.4 &   54.9 &   35.0 &   24.8 &    3.34 &    2.22 &    0.77 &    0.60 \\  
0.25E-09 &  0.000644 &  667.9 &  458.3 &  362.0 &  198.2 &  144.0 &   86.6 &   55.1 &   35.1 &   24.8 &    3.35 &    2.22 &    0.77 &    0.60 \\  
0.50E-09 &  0.001287 &  667.9 &  458.3 &  362.0 &  198.3 &  144.2 &   86.9 &   55.5 &   35.3 &   24.9 &    3.36 &    2.23 &    0.77 &    0.61 \\  
0.10E-08 &  0.002575 &  667.9 &  458.3 &  362.1 &  198.4 &  144.6 &   87.5 &   56.3 &   35.8 &   25.2 &    3.39 &    2.26 &    0.78 &    0.61 \\  
0.25E-08 &  0.006435 &  667.9 &  458.4 &  362.2 &  198.8 &  145.9 &   89.5 &   58.6 &   37.1 &   25.9 &    3.48 &    2.32 &    0.80 &    0.63 \\  
0.50E-08 &  0.012865 &  667.9 &  458.6 &  362.5 &  199.5 &  148.0 &   92.6 &   62.4 &   39.3 &   27.1 &    3.62 &    2.42 &    0.84 &    0.66 \\  
0.10E-07 &  0.025708 &  668.0 &  458.9 &  362.9 &  200.9 &  152.3 &   99.0 &   70.0 &   43.7 &   29.4 &    3.91 &    2.63 &    0.92 &    0.72 \\
\hline
\end{tabular}
\label{tab1}
\end{table*}

\begin{table*}
\caption{ASTRO-F fluxes for a O-rich AGB star, $T_{\rm eff}$ = 3850 K, 60\% Silicate + 40\% AlOx.}
\setlength{\tabcolsep}{1.15mm}
\begin{tabular}{rrrrrrrrrrrrrrr} \hline
\mdot\   & $\tau$    & IRC\_N2 & IRC\_N3 & IRC\_N4 & S7  &   S9W  &   S11  &  L15   &  L18W  &  L24   &  N60  & WIDE-S  & WIDE-L  & N160  \\ \hline

0.10E-09 &  0.000145 &  360.6 &  198.1 &  115.7 &   54.8 &   38.5 &   21.8 &   12.5 &    7.9 &    5.5 &    0.72 &    0.48 &    0.16 &    0.13 \\  
0.25E-09 &  0.000362 &  360.7 &  198.2 &  115.8 &   54.9 &   38.9 &   22.4 &   12.9 &    8.2 &    5.7 &    0.73 &    0.49 &    0.17 &    0.13 \\  
0.50E-09 &  0.000725 &  360.9 &  198.4 &  116.1 &   55.1 &   39.7 &   23.3 &   13.5 &    8.5 &    5.9 &    0.76 &    0.50 &    0.17 &    0.14 \\  
0.10E-08 &  0.001449 &  361.2 &  198.8 &  116.5 &   55.6 &   41.3 &   25.0 &   14.7 &    9.3 &    6.4 &    0.80 &    0.54 &    0.19 &    0.15 \\  
0.25E-08 &  0.003621 &  362.2 &  200.0 &  117.8 &   56.9 &   46.0 &   30.3 &   18.3 &   11.6 &    7.9 &    0.95 &    0.64 &    0.23 &    0.18 \\  
0.50E-08 &  0.007239 &  363.9 &  201.9 &  120.0 &   59.1 &   53.8 &   39.0 &   24.3 &   15.4 &   10.4 &    1.18 &    0.81 &    0.29 &    0.23 \\  
0.10E-07 &  0.014465 &  367.1 &  205.8 &  124.4 &   63.6 &   69.2 &   56.3 &   36.1 &   23.0 &   15.4 &    1.65 &    1.14 &    0.42 &    0.33 \\
\hline
\end{tabular}
\label{tab1}
\end{table*}

\begin{table*}
\caption{ASTRO-F fluxes for a O-rich AGB star, $T_{\rm eff}$ = 3297 K, 60\% Silicate + 40\% AlOx.}
\setlength{\tabcolsep}{1.15mm}
\begin{tabular}{rrrrrrrrrrrrrrr} \hline
\mdot\   & $\tau$    & IRC\_N2 & IRC\_N3 & IRC\_N4 & S7  &   S9W  &   S11  &  L15   &  L18W  &  L24  &  N60  & WIDE-S  & WIDE-L  & N160   \\ \hline

0.10E-09 &  0.000177 &  525.1 &  294.4 &  185.1 &   89.3 &   63.3 &   36.2 &   21.1 &   13.4 &    9.4 &    1.23 &    0.82 &    0.28 &    0.22 \\  
0.25E-09 &  0.000443 &  525.1 &  294.5 &  185.2 &   89.4 &   63.6 &   36.7 &   21.4 &   13.6 &    9.5 &    1.25 &    0.83 &    0.29 &    0.22 \\  
0.50E-09 &  0.000886 &  525.2 &  294.7 &  185.4 &   89.6 &   64.3 &   37.4 &   21.9 &   13.9 &    9.7 &    1.27 &    0.84 &    0.29 &    0.23 \\  
0.10E-08 &  0.001772 &  525.4 &  294.9 &  185.7 &   90.0 &   65.5 &   38.8 &   22.9 &   14.5 &   10.1 &    1.30 &    0.87 &    0.30 &    0.24 \\  
0.25E-08 &  0.004427 &  525.9 &  295.8 &  186.7 &   91.0 &   69.3 &   43.0 &   25.8 &   16.3 &   11.3 &    1.42 &    0.95 &    0.33 &    0.26 \\  
0.50E-08 &  0.008850 &  526.8 &  297.2 &  188.4 &   92.8 &   75.5 &   50.0 &   30.6 &   19.4 &   13.3 &    1.61 &    1.09 &    0.39 &    0.30 \\  
0.10E-07 &  0.017680 &  528.4 &  299.9 &  191.8 &   96.3 &   87.8 &   63.8 &   40.1 &   25.6 &   17.3 &    1.99 &    1.36 &    0.49 &    0.38 \\
\hline
\end{tabular}
\label{tab1}
\end{table*}

\begin{table*}
\caption{ASTRO-F fluxes for a O-rich AGB star, $T_{\rm eff}$ = 2500 K, 60\% Silicate + 40\% AlOx.}
\setlength{\tabcolsep}{1.15mm}
\begin{tabular}{rrrrrrrrrrrrrrr} \hline
\mdot\   & $\tau$    & IRC\_N2 & IRC\_N3 & IRC\_N4 & S7  &   S9W  &   S11  &  L15   &  L18W  &  L24   &  N60  & WIDE-S  & WIDE-L  & N160  \\ \hline

0.10E-09 &  0.000189 &  667.9 &  458.3 &  362.0 &  198.2 &  144.0 &   86.5 &   54.9 &   35.0 &   24.8 &    3.34 &    2.22 &    0.77 &    0.60 \\  
0.25E-09 &  0.000473 &  668.0 &  458.4 &  362.1 &  198.3 &  144.3 &   86.9 &   55.2 &   35.2 &   24.9 &    3.35 &    2.23 &    0.77 &    0.60 \\  
0.50E-09 &  0.000945 &  668.0 &  458.5 &  362.2 &  198.4 &  144.8 &   87.5 &   55.6 &   35.4 &   25.1 &    3.37 &    2.24 &    0.77 &    0.61 \\  
0.10E-08 &  0.001890 &  668.1 &  458.7 &  362.5 &  198.7 &  145.9 &   88.7 &   56.4 &   36.0 &   25.4 &    3.40 &    2.26 &    0.78 &    0.62 \\  
0.25E-08 &  0.004724 &  668.3 &  459.2 &  363.3 &  199.6 &  149.1 &   92.3 &   58.9 &   37.6 &   26.5 &    3.51 &    2.34 &    0.81 &    0.64 \\  
0.50E-08 &  0.009443 &  668.6 &  460.2 &  364.6 &  201.1 &  154.3 &   98.2 &   63.1 &   40.3 &   28.3 &    3.68 &    2.46 &    0.86 &    0.68 \\  
0.10E-07 &  0.018867 &  669.3 &  462.1 &  367.2 &  204.0 &  164.8 &  110.1 &   71.5 &   45.7 &   31.8 &    4.02 &    2.71 &    0.96 &    0.75 \\
\hline
\end{tabular}
\label{tab1}
\end{table*}

\begin{table*}
\caption{ASTRO-F fluxes for a O-rich AGB star, $T_{\rm eff}$ = 3850 K, 100\% Silicate.}
\setlength{\tabcolsep}{1.15mm}
\begin{tabular}{rrrrrrrrrrrrrrr} \hline
\mdot\   & $\tau$    & IRC\_N2 & IRC\_N3 & IRC\_N4 & S7  &   S9W  &   S11  &  L15   &  L18W  &  L24   &  N60  & WIDE-S  & WIDE-L  & N160  \\ \hline

0.10E-09 &  0.000113 &  360.6 &  198.0 &  115.7 &   54.8 &   38.4 &   21.8 &   12.4 &    7.9 &    5.5 &    0.72 &    0.48 &    0.16 &    0.13 \\  
0.25E-09 &  0.000283 &  360.6 &  198.1 &  115.7 &   54.9 &   38.8 &   22.2 &   12.6 &    8.0 &    5.6 &    0.73 &    0.48 &    0.17 &    0.13 \\  
0.50E-09 &  0.000566 &  360.6 &  198.2 &  115.9 &   55.0 &   39.5 &   22.9 &   13.0 &    8.3 &    5.9 &    0.75 &    0.50 &    0.17 &    0.14 \\  
0.10E-08 &  0.001132 &  360.7 &  198.3 &  116.1 &   55.3 &   40.9 &   24.3 &   13.7 &    8.8 &    6.3 &    0.78 &    0.52 &    0.18 &    0.14 \\  
0.25E-08 &  0.002830 &  360.8 &  198.8 &  116.9 &   56.3 &   45.1 &   28.6 &   15.7 &   10.4 &    7.6 &    0.89 &    0.60 &    0.21 &    0.17 \\  
0.50E-08 &  0.005659 &  361.0 &  199.5 &  118.2 &   57.8 &   52.0 &   35.6 &   19.1 &   13.0 &    9.8 &    1.07 &    0.73 &    0.26 &    0.21 \\  
0.10E-07 &  0.011312 &  361.5 &  201.1 &  120.8 &   60.9 &   65.8 &   49.7 &   25.8 &   18.2 &   14.2 &    1.44 &    0.98 &    0.37 &    0.29 \\
\hline
\end{tabular}
\label{tab1}
\end{table*}

\begin{table*}
\caption{ASTRO-F fluxes for a O-rich AGB star, $T_{\rm eff}$ = 3297 K, 100\% Silicate.}
\setlength{\tabcolsep}{1.15mm}
\begin{tabular}{rrrrrrrrrrrrrrr} \hline
\mdot\   & $\tau$    & IRC\_N2 & IRC\_N3 & IRC\_N4 & S7  &   S9W  &   S11  &  L15   &  L18W  &  L24   &  N60  & WIDE-S  & WIDE-L  & N160  \\ \hline

0.10E-09 &  0.000134 &  525.0 &  294.4 &  185.1 &   89.3 &   63.2 &   36.2 &   21.0 &   13.3 &    9.4 &    1.23 &    0.82 &    0.28 &    0.22 \\  
0.25E-09 &  0.000334 &  525.0 &  294.4 &  185.1 &   89.4 &   63.6 &   36.6 &   21.2 &   13.5 &    9.5 &    1.24 &    0.82 &    0.28 &    0.22 \\  
0.50E-09 &  0.000668 &  525.0 &  294.5 &  185.2 &   89.5 &   64.2 &   37.2 &   21.5 &   13.7 &    9.7 &    1.26 &    0.84 &    0.29 &    0.23 \\  
0.10E-08 &  0.001337 &  525.0 &  294.6 &  185.4 &   89.8 &   65.3 &   38.3 &   22.1 &   14.1 &   10.0 &    1.29 &    0.86 &    0.30 &    0.23 \\  
0.25E-08 &  0.003342 &  525.0 &  294.9 &  186.1 &   90.6 &   68.8 &   41.9 &   23.8 &   15.5 &   11.2 &    1.38 &    0.92 &    0.32 &    0.26 \\  
0.50E-08 &  0.006681 &  524.9 &  295.4 &  187.1 &   91.9 &   74.6 &   47.8 &   26.6 &   17.7 &   13.0 &    1.53 &    1.03 &    0.37 &    0.29 \\  
0.10E-07 &  0.013355 &  524.8 &  296.5 &  189.2 &   94.5 &   86.2 &   59.6 &   32.3 &   22.1 &   16.7 &    1.84 &    1.25 &    0.45 &    0.36 \\
\hline
\end{tabular}
\label{tab1}
\end{table*}

\begin{table*}
\caption{ASTRO-F fluxes for a O-rich AGB star, $T_{\rm eff}$ = 2500 K, 100\% Silicate.}
\setlength{\tabcolsep}{1.15mm}
\begin{tabular}{rrrrrrrrrrrrrrr} \hline
\mdot\   & $\tau$    & IRC\_N2 & IRC\_N3 & IRC\_N4 & S7  &   S9W  &   S11  &  L15   &  L18W  &  L24   &  N60  & WIDE-S  & WIDE-L  & N160  \\ \hline

0.10E-09 &  0.000144 &  667.9 &  458.3 &  362.0 &  198.2 &  144.0 &   86.5 &   54.8 &   35.0 &   24.8 &    3.34 &    2.22 &    0.77 &    0.60 \\  
0.25E-09 &  0.000361 &  667.9 &  458.3 &  362.1 &  198.3 &  144.3 &   86.8 &   55.0 &   35.1 &   24.9 &    3.35 &    2.22 &    0.77 &    0.60 \\  
0.50E-09 &  0.000722 &  667.9 &  458.4 &  362.1 &  198.4 &  144.8 &   87.3 &   55.3 &   35.3 &   25.1 &    3.36 &    2.23 &    0.77 &    0.61 \\  
0.10E-08 &  0.001445 &  667.8 &  458.4 &  362.3 &  198.6 &  145.8 &   88.4 &   55.8 &   35.7 &   25.4 &    3.39 &    2.25 &    0.78 &    0.61 \\  
0.25E-08 &  0.003611 &  667.6 &  458.6 &  362.8 &  199.3 &  149.0 &   91.6 &   57.3 &   36.9 &   26.4 &    3.47 &    2.31 &    0.81 &    0.63 \\  
0.50E-08 &  0.007219 &  667.4 &  459.0 &  363.7 &  200.4 &  154.2 &   96.9 &   59.9 &   38.9 &   28.1 &    3.62 &    2.41 &    0.85 &    0.67 \\  
0.10E-07 &  0.014430 &  666.8 &  459.6 &  365.3 &  202.7 &  164.5 &  107.5 &   65.1 &   42.9 &   31.5 &    3.90 &    2.61 &    0.93 &    0.73 \\
\hline
\end{tabular}
\label{tab1}
\end{table*}



\begin{table*}
\caption{Optical, NIR, IRAC and MIPS magnitudes and fluxes for a C-rich post-AGB star, $T_{\rm eff}$ = 2650 K, 100\% AMC.}
\begin{tabular}{rrrrrrrrrrrrrr} \hline
\small

$T_{\rm c}$   & $\tau$    &  $V$  &  $I$  &  $J$  &  $H$  &  $K$  &    3.6 &    4.5 &    5.8 &    8.0 &   24    &   70    & 160  \\ \hline

0.12E+04 &  1.046662 & 42.61 & 31.70 & 21.35 & 15.91 & 12.05 &  329.8 &  871.1 & 1599.8 & 2429.9 & 1568.50 &  372.92 & 134.401 \\  
0.11E+04 &  0.903924 & 38.97 & 29.06 & 20.22 & 15.38 & 11.83 &  344.1 &  886.1 & 1601.1 & 2420.3 & 1547.85 &  371.27 & 134.476 \\  
0.10E+04 &  0.763922 & 35.32 & 26.32 & 18.71 & 14.58 & 11.50 &  360.1 &  899.1 & 1595.3 & 2405.1 & 1530.56 &  371.33 & 135.275 \\  
0.90E+03 &  0.628735 & 31.73 & 23.65 & 17.08 & 13.59 & 11.07 &  375.2 &  905.4 & 1577.7 & 2384.3 & 1519.30 &  373.95 & 137.131 \\  
0.80E+03 &  0.496372 & 28.15 & 21.01 & 15.42 & 12.48 & 10.50 &  389.2 &  902.4 & 1543.5 & 2349.8 & 1507.28 &  377.22 & 139.409 \\  
0.70E+03 &  0.373292 & 24.74 & 18.55 & 13.86 & 11.39 &  9.87 &  388.1 &  868.4 & 1472.2 & 2295.2 & 1506.54 &  384.40 & 143.332 \\  
0.60E+03 &  0.264308 & 21.65 & 16.36 & 12.47 & 10.39 &  9.24 &  352.6 &  765.6 & 1318.2 & 2187.6 & 1545.88 &  405.25 & 152.731 \\
\hline
\end{tabular}
\label{tab1}
\end{table*}

\begin{table*}
\caption{Optical, NIR, IRAC and MIPS magnitudes and fluxes for a C-rich post-AGB star, $T_{\rm eff}$ = 3600 K, 100\% AMC.}
\begin{tabular}{rrrrrrrrrrrrrr} \hline
\small

\mdot\   & $\tau$    &  $V$  &  $I$  &  $J$  &  $H$  &  $K$  &    3.6 &    4.5 &    5.8 &    8.0 &   24    &   70    & 160  \\ \hline

0.12E+04 &  0.980290 & 39.63 & 30.04 & 20.91 & 15.76 & 11.98 &  339.3 &  880.1 & 1607.4 & 2435.2 & 1549.74 &  370.21 & 133.849 \\  
0.11E+04 &  0.839955 & 35.97 & 27.32 & 19.62 & 15.18 & 11.76 &  355.3 &  893.8 & 1606.3 & 2424.3 & 1529.45 &  369.09 & 134.146 \\  
0.10E+04 &  0.702712 & 32.30 & 24.61 & 18.04 & 14.36 & 11.44 &  373.8 &  904.9 & 1597.5 & 2406.7 & 1512.62 &  369.83 & 135.247 \\  
0.90E+03 &  0.570039 & 28.68 & 21.97 & 16.41 & 13.37 & 11.02 &  394.5 &  912.1 & 1579.2 & 2381.3 & 1495.84 &  371.46 & 136.809 \\  
0.80E+03 &  0.445435 & 25.20 & 19.47 & 14.83 & 12.33 & 10.54 &  409.4 &  904.1 & 1542.9 & 2348.4 & 1486.45 &  375.95 & 139.613 \\  
0.70E+03 &  0.331322 & 21.94 & 17.17 & 13.37 & 11.31 & 10.01 &  406.9 &  864.1 & 1473.9 & 2301.8 & 1488.51 &  383.98 & 143.922 \\  
0.60E+03 &  0.232149 & 19.04 & 15.16 & 12.09 & 10.40 &  9.48 &  364.6 &  751.6 & 1326.4 & 2215.1 & 1535.93 &  406.23 & 153.864 \\
\hline
\end{tabular}
\label{tab1}
\end{table*}

\begin{table*}
\caption{Optical, NIR, IRAC and MIPS magnitudes and fluxes for a O-rich post-AGB star, $T_{\rm eff}$ = 3850 K, 100\% Silicate.}
\begin{tabular}{rrrrrrrrrrrrrr} \hline
\small

$T_{\rm c}$   & $\tau$    &  $V$  &  $I$  &  $J$  &  $H$  &  $K$  &    3.6 &    4.5 &    5.8 &    8.0 &   24    &   70    & 160  \\ \hline

0.10E+04 & 15.022061 & 33.24 & 24.39 & 17.93 & 14.42 & 11.70 &  291.9 &  695.8 & 1249.6 & 1717.1 & 4205.81 &  675.25 & 224.784 \\  
0.90E+03 & 12.476484 & 29.83 & 22.04 & 16.50 & 13.56 & 11.40 &  290.4 &  678.1 & 1212.6 & 1692.6 & 4243.28 &  675.71 & 227.586 \\  
0.80E+03 & 10.099082 & 26.58 & 19.81 & 15.10 & 12.62 & 11.01 &  279.9 &  641.3 & 1149.2 & 1656.0 & 4313.57 &  678.77 & 231.449 \\  
0.70E+03 &  7.949813 & 23.57 & 17.79 & 13.83 & 11.72 & 10.56 &  250.3 &  566.7 & 1036.3 & 1595.3 & 4439.93 &  685.39 & 236.700 \\  
0.60E+03 &  6.077328 & 20.89 & 16.02 & 12.72 & 10.90 & 10.07 &  198.4 &  439.9 &  846.2 & 1477.4 & 4639.07 &  699.86 & 245.028 \\  
0.50E+03 &  4.486949 & 18.58 & 14.50 & 11.77 & 10.20 &  9.58 &  142.3 &  277.8 &  577.4 & 1248.2 & 4865.82 &  724.29 & 257.750 \\  
0.40E+03 &  3.125848 & 16.57 & 13.20 & 10.96 &  9.60 &  9.14 &  114.2 &  141.7 &  295.8 &  878.4 & 4966.26 &  754.68 & 274.670 \\
\hline
\end{tabular}
\label{tab1}
\end{table*}

\begin{table*}
\caption{Optical, NIR, IRAC and MIPS magnitudes and fluxes for a O-rich post-AGB star, $T_{\rm eff}$ = 3297 K, 100\% Silicate.}
\begin{tabular}{rrrrrrrrrrrrrr} \hline
\small

$T_{\rm c}$   & $\tau$    &  $V$  &  $I$  &  $J$  &  $H$  &  $K$  &    3.6 &    4.5 &    5.8 &    8.0 &   24    &   70    & 160  \\ \hline

0.10E+04 & 16.290331 & 37.43 & 26.26 & 18.42 & 14.58 & 11.74 &  289.7 &  695.2 & 1255.3 & 1719.6 & 4193.92 &  678.37 & 224.747 \\  
0.90E+03 & 13.652185 & 34.02 & 23.87 & 16.97 & 13.71 & 11.42 &  288.9 &  678.8 & 1221.4 & 1698.2 & 4221.60 &  675.79 & 226.391 \\  
0.80E+03 & 11.150371 & 30.67 & 21.56 & 15.51 & 12.73 & 10.98 &  281.5 &  645.2 & 1163.3 & 1663.5 & 4282.63 &  678.58 & 230.116 \\  
0.70E+03 &  8.841756 & 27.48 & 19.42 & 14.14 & 11.75 & 10.44 &  258.6 &  575.9 & 1055.5 & 1601.6 & 4394.52 &  684.52 & 235.105 \\  
0.60E+03 &  6.794093 & 24.57 & 17.52 & 12.93 & 10.86 &  9.87 &  217.1 &  455.6 &  868.9 & 1478.3 & 4572.09 &  697.31 & 242.847 \\  
0.50E+03 &  5.037727 & 22.01 & 15.87 & 11.89 & 10.08 &  9.31 &  173.5 &  299.9 &  600.9 & 1240.6 & 4767.33 &  718.30 & 254.375 \\  
0.40E+03 &  3.533377 & 19.76 & 14.46 & 10.99 &  9.42 &  8.81 &  159.5 &  170.3 &  319.5 &  864.0 & 4816.21 &  741.63 & 268.796 \\
\hline
\end{tabular}
\label{tab1}
\end{table*}

\begin{table*}
\caption{Optical, NIR, IRAC and MIPS magnitudes and fluxes for a O-rich post-AGB star, $T_{\rm eff}$ = 2500 K, 100\% Silicate.}
\begin{tabular}{rrrrrrrrrrrrrr} \hline
\small

$T_{\rm c}$   & $\tau$    &  $V$  &  $I$  &  $J$  &  $H$  &  $K$  &    3.6 &    4.5 &    5.8 &    8.0 &   24    &   70    & 160  \\ \hline

0.10E+04 & 16.987080 & 40.68 & 28.28 & 18.62 & 14.79 & 11.79 &  289.4 &  698.3 & 1264.4 & 1726.4 & 4206.51 &  685.39 & 226.565 \\  
0.90E+03 & 14.282159 & 37.29 & 25.88 & 17.13 & 13.95 & 11.46 &  291.0 &  685.1 & 1232.9 & 1705.2 & 4221.75 &  679.07 & 226.872 \\  
0.80E+03 & 11.712989 & 33.95 & 23.54 & 15.63 & 12.96 & 10.99 &  288.7 &  657.6 & 1179.7 & 1669.2 & 4270.69 &  680.71 & 230.207 \\  
0.70E+03 &  9.322069 & 30.75 & 21.35 & 14.22 & 11.95 & 10.42 &  275.9 &  599.1 & 1080.0 & 1604.8 & 4362.60 &  684.98 & 234.649 \\  
0.60E+03 &  7.182675 & 27.78 & 19.39 & 12.95 & 11.03 &  9.80 &  250.5 &  496.1 &  906.6 & 1479.8 & 4509.55 &  694.86 & 241.422 \\  
0.50E+03 &  5.337032 & 25.14 & 17.69 & 11.86 & 10.22 &  9.22 &  227.6 &  362.4 &  656.7 & 1245.8 & 4666.47 &  711.31 & 251.375 \\  
0.40E+03 &  3.754286 & 22.80 & 16.22 & 10.92 &  9.52 &  8.70 &  237.0 &  256.4 &  395.9 &  881.5 & 4676.07 &  728.42 & 263.522 \\
\hline
\end{tabular}
\label{tab1}
\end{table*}


\begin{table*}
\caption{ASTRO-F fluxes for a C-rich post-AGB star, $T_{\rm eff}$ = 2650 K, 100\% AMC.}
\setlength{\tabcolsep}{1.15mm}
\begin{tabular}{rrrrrrrrrrrrrrr} \hline

$T_{\rm c}$   & $\tau$    & IRC\_N2 & IRC\_N3 & IRC\_N4 & S7  &   S9W  &   S11  &  L15   &  L18W  &  L24   &  N60  & WIDE-S  & WIDE-L  & N160  \\ \hline

0.12E+04 &  1.046662 &   33.8 &  200.9 &  835.1 & 2475.2 & 2863.3 & 2473.0 & 2368.5 & 1754.2 & 1588.1 &  417.23 &  318.25 &  142.47 &  117.77 \\  
0.11E+04 &  0.903924 &   38.3 &  211.6 &  850.2 & 2468.3 & 2842.6 & 2446.5 & 2331.5 & 1728.1 & 1567.0 &  415.12 &  317.18 &  142.46 &  117.90 \\  
0.10E+04 &  0.763922 &   45.4 &  224.1 &  863.9 & 2454.4 & 2816.5 & 2416.9 & 2295.3 & 1703.8 & 1549.3 &  414.86 &  317.64 &  143.21 &  118.66 \\  
0.90E+03 &  0.628735 &   56.5 &  237.2 &  871.3 & 2432.2 & 2785.9 & 2386.2 & 2262.6 & 1683.8 & 1537.6 &  417.38 &  320.38 &  145.08 &  120.35 \\  
0.80E+03 &  0.496372 &   76.1 &  251.4 &  870.4 & 2393.6 & 2741.1 & 2345.6 & 2224.5 & 1661.0 & 1525.0 &  420.56 &  323.82 &  147.39 &  122.41 \\  
0.70E+03 &  0.373292 &  108.6 &  258.0 &  839.7 & 2326.5 & 2681.9 & 2302.2 & 2195.3 & 1647.4 & 1523.7 &  428.02 &  330.72 &  151.42 &  125.92 \\  
0.60E+03 &  0.264308 &  160.9 &  246.4 &  742.7 & 2187.5 & 2587.5 & 2256.6 & 2201.8 & 1667.9 & 1562.5 &  450.56 &  349.61 &  161.21 &  134.26 \\
\hline
\end{tabular}
\label{tab1}
\end{table*}

\begin{table*}
\caption{ASTRO-F fluxes for a C-rich post-AGB star, $T_{\rm eff}$ = 3600 K, 100\% AMC.}
\setlength{\tabcolsep}{1.15mm}
\begin{tabular}{rrrrrrrrrrrrrrr} \hline

$T_{\rm c}$  & $\tau$    & IRC\_N2 & IRC\_N3 & IRC\_N4 & S7  &   S9W  &   S11  &  L15   &  L18W  &  L24   &  N60  & WIDE-S  & WIDE-L  & N160  \\ \hline

0.12E+04 &  0.980290 &   35.5 &  208.1 &  844.7 & 2483.2 & 2862.8 & 2466.3 & 2352.1 & 1738.7 & 1569.2 &  414.03 &  316.18 &  141.83 &  117.33 \\  
0.11E+04 &  0.839955 &   40.3 &  220.4 &  859.0 & 2474.7 & 2840.4 & 2438.3 & 2314.6 & 1712.5 & 1548.5 &  412.49 &  315.58 &  142.06 &  117.65 \\  
0.10E+04 &  0.702712 &   47.7 &  235.6 &  871.4 & 2458.3 & 2811.3 & 2406.4 & 2277.7 & 1688.1 & 1531.2 &  412.95 &  316.67 &  143.14 &  118.68 \\  
0.90E+03 &  0.570039 &   59.0 &  253.7 &  880.7 & 2432.6 & 2773.3 & 2367.6 & 2236.7 & 1661.9 & 1513.8 &  414.34 &  318.62 &  144.69 &  120.11 \\  
0.80E+03 &  0.445435 &   75.1 &  269.7 &  875.5 & 2395.3 & 2730.1 & 2328.5 & 2201.3 & 1641.5 & 1503.9 &  418.85 &  323.16 &  147.55 &  122.63 \\  
0.70E+03 &  0.331322 &   98.2 &  276.1 &  839.2 & 2336.6 & 2679.2 & 2291.4 & 2177.1 & 1631.3 & 1505.4 &  427.23 &  330.83 &  151.99 &  126.48 \\  
0.60E+03 &  0.232149 &  130.4 &  259.5 &  732.7 & 2218.1 & 2608.7 & 2265.6 & 2199.4 & 1662.4 & 1552.4 &  451.35 &  350.92 &  162.34 &  135.29 \\
\hline
\end{tabular}
\label{tab1}
\end{table*}

\begin{table*}
\caption{ASTRO-F fluxes for a O-rich post-AGB star, $T_{\rm eff}$ = 3850 K, 100\% Silicate.}
\setlength{\tabcolsep}{1.15mm}
\begin{tabular}{rrrrrrrrrrrrrrr} \hline

$T_{\rm c}$   & $\tau$    & IRC\_N2 & IRC\_N3 & IRC\_N4 & S7  &   S9W  &   S11  &  L15   &  L18W  &  L24   &  N60  & WIDE-S  & WIDE-L  & N160  \\ \hline

0.10E+04 & 15.022061 &   36.4 &  183.4 &  671.0 & 1962.5 & 1494.7 & 1364.0 & 3761.2 & 3551.3 & 4221.2 &  776.60 &  556.97 &  235.97 &  197.87 \\  
0.90E+03 & 12.476484 &   41.3 &  184.8 &  655.0 & 1913.1 & 1490.7 & 1389.8 & 3814.2 & 3593.5 & 4260.0 &  776.43 &  558.33 &  238.47 &  200.63 \\  
0.80E+03 & 10.099082 &   48.3 &  181.1 &  620.4 & 1834.8 & 1489.2 & 1438.1 & 3909.7 & 3671.5 & 4332.7 &  779.07 &  562.07 &  242.10 &  204.31 \\  
0.70E+03 &  7.949813 &   58.3 &  166.1 &  549.3 & 1703.1 & 1499.1 & 1534.9 & 4080.8 & 3813.9 & 4463.9 &  785.59 &  568.99 &  247.21 &  209.19 \\  
0.60E+03 &  6.077328 &   74.3 &  139.0 &  428.1 & 1478.3 & 1516.9 & 1691.9 & 4330.6 & 4033.0 & 4670.9 &  800.82 &  582.76 &  255.55 &  216.78 \\  
0.50E+03 &  4.486949 &  102.2 &  113.6 &  274.4 & 1125.6 & 1509.2 & 1865.4 & 4560.9 & 4265.5 & 4907.1 &  826.85 &  605.42 &  268.44 &  228.26 \\  
0.40E+03 &  3.125848 &  145.9 &  110.0 &  147.3 &  676.5 & 1399.4 & 1908.2 & 4524.3 & 4323.7 & 5013.8 &  858.44 &  634.31 &  285.55 &  243.52 \\
\hline
\end{tabular}
\label{tab1}
\end{table*}

\begin{table*}
\caption{ASTRO-F fluxes for a O-rich post-AGB star, $T_{\rm eff}$ = 3297 K, 100\% Silicate.}
\setlength{\tabcolsep}{1.15mm}
\begin{tabular}{rrrrrrrrrrrrrrr} \hline

$T_{\rm c}$   & $\tau$    & IRC\_N2 & IRC\_N3 & IRC\_N4 & S7  &   S9W  &   S11  &  L15   &  L18W  &  L24   &  N60  & WIDE-S  & WIDE-L  & N160  \\ \hline

0.10E+04 & 16.290331 &   34.9 &  181.0 &  670.4 & 1973.8 & 1487.0 & 1345.6 & 3734.3 & 3532.2 & 4208.4 &  780.38 &  559.18 &  236.12 &  197.71 \\  
0.90E+03 & 13.652185 &   39.6 &  182.6 &  655.7 & 1928.9 & 1487.2 & 1373.7 & 3782.9 & 3568.7 & 4237.5 &  776.84 &  557.94 &  237.41 &  199.44 \\  
0.80E+03 & 11.150371 &   47.6 &  180.9 &  624.7 & 1858.1 & 1482.8 & 1412.8 & 3863.5 & 3634.9 & 4300.5 &  779.24 &  561.36 &  240.88 &  203.01 \\  
0.70E+03 &  8.841756 &   61.1 &  171.1 &  559.5 & 1732.1 & 1483.4 & 1492.1 & 4010.8 & 3758.2 & 4416.4 &  785.04 &  567.63 &  245.70 &  207.68 \\  
0.60E+03 &  6.794093 &   84.7 &  152.6 &  446.0 & 1508.7 & 1481.6 & 1621.1 & 4228.5 & 3950.6 & 4600.7 &  798.37 &  579.96 &  253.42 &  214.76 \\  
0.50E+03 &  5.037727 &  125.3 &  138.0 &  300.2 & 1151.6 & 1442.1 & 1756.5 & 4420.8 & 4148.5 & 4804.3 &  820.46 &  599.75 &  265.05 &  225.20 \\  
0.40E+03 &  3.533377 &  188.2 &  147.2 &  181.2 &  695.2 & 1294.9 & 1757.9 & 4338.0 & 4159.6 & 4858.4 &  843.97 &  622.74 &  279.56 &  238.25 \\
\hline
\end{tabular}
\label{tab1}
\end{table*}

\begin{table*}
\caption{ASTRO-F fluxes for a O-rich post-AGB star, $T_{\rm eff}$ = 2500 K, 100\% Silicate.}
\setlength{\tabcolsep}{1.15mm}
\begin{tabular}{rrrrrrrrrrrrrrr} \hline

$T_{\rm c}$   & $\tau$    & IRC\_N2 & IRC\_N3 & IRC\_N4 & S7  &   S9W  &   S11  &  L15   &  L18W  &  L24   &  N60  & WIDE-S  & WIDE-L  & N160  \\ \hline

0.10E+04 & 16.987080 &   34.3 &  180.4 &  673.5 & 1987.8 & 1485.7 & 1337.2 & 3731.9 & 3535.2 & 4220.3 &  788.45 &  564.80 &  238.13 &  199.25 \\  
0.90E+03 & 14.282159 &   39.5 &  183.7 &  662.1 & 1943.4 & 1488.5 & 1367.4 & 3774.9 & 3564.6 & 4237.3 &  780.69 &  560.43 &  238.03 &  199.79 \\  
0.80E+03 & 11.712989 &   49.0 &  186.3 &  637.5 & 1875.0 & 1480.6 & 1399.1 & 3840.6 & 3618.2 & 4287.9 &  781.80 &  562.87 &  241.08 &  203.02 \\  
0.70E+03 &  9.322069 &   66.3 &  184.8 &  584.0 & 1752.5 & 1473.0 & 1463.7 & 3962.1 & 3720.1 & 4383.2 &  785.69 &  567.74 &  245.32 &  207.22 \\  
0.60E+03 &  7.182675 &   97.2 &  180.0 &  488.8 & 1535.8 & 1457.8 & 1569.2 & 4142.5 & 3880.0 & 4536.0 &  795.69 &  577.64 &  252.00 &  213.45 \\  
0.50E+03 &  5.337032 &  149.4 &  183.8 &  366.2 & 1190.6 & 1402.8 & 1677.2 & 4295.1 & 4040.5 & 4700.4 &  812.58 &  593.64 &  261.98 &  222.50 \\  
0.40E+03 &  3.754286 &  228.9 &  214.9 &  272.7 &  751.0 & 1246.1 & 1659.5 & 4183.8 & 4018.9 & 4714.6 &  829.02 &  611.39 &  274.12 &  233.55 \\
\hline
\end{tabular}
\label{tab1}
\end{table*}

\end{document}